\documentclass[12pt]{article}
\pdfoutput=1
\usepackage[a4paper]{geometry}
\usepackage{jheppub, amsmath,amssymb,amsfonts,amsxtra, mathrsfs, makeidx,graphics,graphicx,amsthm,epsfig, bm,longtable,float, color,tikz,mathtools,xfrac,footnote,rotating, lscape, makecell, environ,mathtools, empheq}

\usepackage{amsmath,amscd,amstext,amsbsy,amsopn,amsthm,amsxtra,upref,amsfonts,amssymb,euscript,latexsym,graphicx,color,braket,slashed,multicol,mdframed,tikz,tikz-3dplot,todonotes,hyperref,pgfplots,mathrsfs,bbm,array,cleveref,graphicx,wrapfig,lscape,youngtab,bbold,float,multirow,longtable,rotating,epstopdf,tikz-cd,parskip}  % Added by Antoine
 
\usepackage[T1]{fontenc} 

\usepackage{mathtools}% http://ctan.org/pkg/mathtools
\usepackage{colortbl}
\usepackage{nicematrix}
\usepackage{diagbox}

\usepackage{subfig}
\usepackage{multirow}
\usepackage{adjustbox}

\usepackage{amsthm}
\usepackage{hyperref}

\usepackage{amsmath}
\usepackage{amsfonts}
\usepackage{commath}
\usepackage[english]{babel}
\usepackage{setspace}
\usepackage{youngtab}

\usepackage{epsfig,amssymb} 
\usepackage{amsmath}

\usepackage{amscd,color}
\usepackage{amsmath,graphicx}
\usepackage{verbatim,booktabs}

\usepackage{latexsym}
\usepackage{amsmath,amsfonts,amssymb,amsthm}
\usepackage{amsmath,amsthm}

\usepackage{xstring}

\DeclareMathOperator{\Pf}{Pf}

\newcommand{\zz}{\mathbb{Z}}

\newcommand{\cc}{\mathbb{C}}

\newcommand{\NN}{\mathcal{N}}

\newcommand{\CB}{CB }

\newlength\mylen

\usepackage{tikz}
\usetikzlibrary{tikzmark}
\usetikzlibrary{math} %needed tikz library
\usetikzlibrary{matrix}
\usetikzlibrary{calc}
\usetikzlibrary{positioning}
\definecolor{azzurro}{RGB}{0,118,186}
\definecolor{rosso}{RGB}{255,0,0}

\usepackage{array}

\graphicspath{ {./images/} }

\definecolor{mandarancio}{RGB}{255,147,0}
\definecolor{azzurro}{RGB}{0,118,186}
\definecolor{terrabruciatadiconcorezzo}{RGB}{181,23,0}
\def\centerarc[#1](#2)(#3:#4:#5)% Syntax: [draw options] (center) (initial angle:final angle:radius)
{ \draw[#1] ($(#2)+({#5*cos(#3)},{#5*sin(#3)})$) arc (#3:#4:#5) }

\def\matSL[#1,#2;#3,#4] 
{	\left( \begin{array}{cc} #1 & #2 \\ #3 & #4 \end{array} \right)}

\def\pf[#1] {\text{Pf} \left( #1 \right)}

\newcommand{\munutable}[6] % Syntax: [theory][r]{(1,1), (-1,1), (1,-1), (-1,-1)}
{
\begin{equation}
\begin{array}{|c|l|l|}
\hline r=#2 & \multicolumn{2}{c|}{#1}\\
\hline \nu \backslash \mu & \multicolumn{1}{c|}{1} & \multicolumn{1}{c|}{-1} \\
\hline 1 &\begin{array}{l} #3 \end{array} &\begin{array}{l} #4 \end{array} \\
\hline-1 & \begin{array}{l} #5 \end{array} & \begin{array}{l} #6 \end{array} \\
\hline
\end{array}
\end{equation}
}

\def\strata(#1){ \mathcal{S}^{(#1)} }
\def\DD(#1,#2){\left\langle #1, #2 \right\rangle}

\def\extensiondiagram{\text{Ext}}

\def\SFoldNotation{2}
\if\SFoldNotation1
	\def\sfold(#1,#2){ #1 S_{#2}\text{-fold} }
	\def\sfolds(#1,#2){ #1 S_{#2}\text{-folds} }
\fi
\if\SFoldNotation2
	\def\sfold(#1,#2){ S_{#2}\text{-fold} \text{ of type }#1}
	\def\sfolds(#1,#2){ S_{#2}\text{-folds}  \text{ of type }#1}
\fi

\def\algstate[#1,#2] { \left| #1 ; \, #2 \right\rangle}
\def\longstate[#1,#2]{\overline{ \left| #1 , #2 \right\rangle}}

\usepackage{booktabs}
\def\singlenode{\begin{tikzpicture} \node[draw,circle,thick] {} (0,0); \end{tikzpicture} }
\def\doublenode{\begin{tikzpicture} \node[draw,double,circle,thick] {} (0,0); \end{tikzpicture} }

\tikzstyle{newarrowstyle1} = []
\tikzstyle{newarrowstyle2} = []
\tikzstyle{colorAnsubgraph} = [color=black]

\tikzstyle{doublecirclestyle}=[double,circle,thick,draw=black, font={\footnotesize},minimum size = 15pt]
\tikzstyle{smalldoublecirclestyle}=[double,circle,thick,draw=black, font={\footnotesize},minimum size = 10pt]

\tikzstyle{singlecirclestyle}=[circle,thick,draw=black, font={\footnotesize},minimum size = 15pt]

\tikzstyle{graycirclestyle} = [dashed,gray,circle,thick,draw,font={\footnotesize},minimum size = 15pt]

\tikzstyle{arrowstyle}=[{draw,very thick}]

\def\arrowedge[#1](#2,#3)(#4){ %\arrowedge[label] (node1, node2) (edge options)
\IfEq{#1}{0}{}
{
	\IfEq{#1}{1} {\path [-,#4] (#2) edge  node {$1$} (#3);}
	{
		\IfEq{#1}{-1}  { \path [-,#4] (#2) edge (#3);}
		{
			\path [->,#4] (#2) edge node {$#1$} (#3);
		}
	}
}
}

\def\arrowedgetwo(#1,#2,#3)(#4){ %\arrowedge[label] (node1, node2) (edge options)
\IfEq{#1}{0}{}
{
	\IfEq{#1}{1} {\path [-,#4] (#2) edge  node {$1$} (#3);}
	{
		\IfEq{#1}{-1}  { \path [-,#4] (#2) edge (#3);}
		{
			\path [->,#4] (#2) edge node {$#1$} (#3);
		}
	}
}
}

\def\diagramKfive[#1](#2,#3,#4,#5,#6){
\vcenter{\hbox{\begin{tikzpicture}[#1]
\begin{scope} [every node/.style=doublecirclestyle]
\node[label=below:#2,double,circle,thick] (A) at (0,0) {};
\node[label=below:#3,double,circle,thick]  (C) at (2,0) {};
\node[label=below:#4,double,circle,thick]  (D) at (4,0) {};
\node[label=below:#5,double,circle,thick]  (E) at (6,0) {};
\node[double,circle,thick] (F) at (3,1.7) {#6};
\end{scope}
\begin{scope}[every node/.style={fill=white}, every edge/.style={draw,very thick,above,right}]
\path [-] (C) edge (D);
\path [-] (A) edge (C);
\path [-] (D) edge (E);
\path [-] (C) edge (F);
\path [->] (F) edge node[right=0.25,above] {$\omega$} (D);
\end{scope}
\end{tikzpicture}}}
}

\def\diagramKsix[#1](#2,#3,#4,#5,#6,#7){
\vcenter{\hbox{\begin{tikzpicture}[#1]
\begin{scope} [every node/.style=doublecirclestyle]
\node[label=below:#2,double,circle,thick] (A) at (0,0) {};
\node[label=below:#3,double,circle,thick]  (C) at (2,0) {};
\node[label=below:#4,double,circle,thick]  (D) at (4,0) {};
\node[label=below:#5,double,circle,thick]  (E) at (6,0) {};
\node[double,circle,thick] (F) at (3,1.7) {#7};
\node[label=below:#6,double,circle,thick]  (G) at (8,0) {};
\end{scope}
\begin{scope}[every node/.style={fill=white}, every edge/.style={draw,very thick,above,right}]
\path [-] (C) edge (D);
\path [-] (A) edge (C);
\path [-] (D) edge (E);
\path [-] (C) edge (F);
\path [-] (E) edge (G);
\path [->] (F) edge node[right=0.25,above] {$\omega$} (D);
\end{scope}
\end{tikzpicture}}}
}

\def\diagramKfivesmall[#1](#2,#3,#4,#5,#6){
\vcenter{\hbox{\begin{tikzpicture}[#1]
\begin{scope} [every node/.style=smalldoublecirclestyle]
\node[label=below:#2,double,circle,thick] (A) at (0,0) {};
\node[label=below:#3,double,circle,thick]  (C) at (2,0) {};
\node[label=below:#4,double,circle,thick]  (D) at (4,0) {};
\node[label=below:#5,double,circle,thick]  (E) at (6,0) {};
\node[double,circle,thick] (F) at (3,1.7) {#6};
\end{scope}
\begin{scope}[every node/.style={fill=white}, every edge/.style={draw,very thick,above,right}]
\path [-] (C) edge (D);
\path [-] (A) edge (C);
\path [-] (D) edge (E);
\path [-] (C) edge (F);
\path [->] (F) edge node[right=0.25,above] {$\omega$} (D);
\end{scope}
\end{tikzpicture}}}
}

\def\diagramKsixsmall[#1](#2,#3,#4,#5,#6,#7){
\vcenter{\hbox{\begin{tikzpicture}[#1]
\begin{scope} [every node/.style=smalldoublecirclestyle]
\node[label=below:#2,double,circle,thick] (A) at (0,0) {};
\node[label=below:#3,double,circle,thick]  (C) at (2,0) {};
\node[label=below:#4,double,circle,thick]  (D) at (4,0) {};
\node[label=below:#5,double,circle,thick]  (E) at (6,0) {};
\node[double,circle,thick] (F) at (3,1.7) {#7};
\node[label=below:#6,double,circle,thick]  (G) at (8,0) {};
\end{scope}
\begin{scope}[every node/.style={fill=white}, every edge/.style={draw,very thick,above,right}]
\path [-] (C) edge (D);
\path [-] (A) edge (C);
\path [-] (D) edge (E);
\path [-] (C) edge (F);
\path [-] (E) edge (G);
\path [->] (F) edge node[right=0.25,above] {$\omega$} (D);
\end{scope}
\end{tikzpicture}}}
}

\def\diagramA[#1](#2)(#3,#4,#5) { % \diagramA[draw options](edge new-1, edge new-2, edge new-3)
\begin{tikzpicture}[#1]
\begin{scope} [every node/.style=doublecirclestyle]
	\node (A) at (0,0) {1};
	\node (B) at (2,0) {2};
	\node (C) at (4,0) {3};
	\node (D) at (6,0) {};
	\node (E) at (8,0) {$n$};
	\ifthenelse{#3=0 \AND #4=0 \AND #5=0}{}{	\node (new) at (2,2) {#2};}
\end{scope}

\begin{scope}[every node/.style={fill=white,circle}, every edge/.style={draw,very thick}]
	\path [-] (A) edge (B);
	\path [-] (B) edge (C);
	\path [dashed] (C) edge (D);
	\path [-] (D) edge (E);
\end{scope}

\begin{scope}[every node/.style={fill=white}, every edge/.style={draw,very thick}]
		\arrowedge[#3](new,A)(bend right,newarrowstyle1)
		\arrowedge[#4](new,B)(newarrowstyle1)
		\arrowedge[#5](new,C)(bend left,newarrowstyle1)
\end{scope}

\end{tikzpicture}
}

\def\diagramAbar[#1](#2)(#3,#4,#5) { % \diagramA[draw options](edge new-1, edge new-2, edge new-3)
\begin{tikzpicture}[#1]
	\node[singlecirclestyle] (A) at (0,0) {1};
\begin{scope} [every node/.style=doublecirclestyle]
	\node (B) at (2,0) {2};
	\node (C) at (4,0) {3};
	\node (D) at (6,0) {};
	\node (E) at (8,0) {$n$};
	\ifthenelse{#3=0 \AND #4=0 \AND #5=0}{}{	\node (new) at (2,2) {#2};}
\end{scope}

\begin{scope}[every node/.style={fill=white,circle}, every edge/.style={draw,very thick}]
	\path [-] (A) edge (B);
	\path [-] (B) edge (C);
	\path [dashed] (C) edge (D);
	\path [-] (D) edge (E);
\end{scope}

\begin{scope}[every node/.style={fill=white}, every edge/.style={draw,very thick}]
		\arrowedge[#3](new,A)(bend right,newarrowstyle1)
		\arrowedge[#4](new,B)(newarrowstyle1)
		\arrowedge[#5](new,C)(bend left,newarrowstyle1)
\end{scope}

\end{tikzpicture}
}

\def\sdiagramD[#1](#2)(#3,#4,#5,#6) { % \diagramD[draw options](edge new-1, edge new-2, edge new-3)
\begin{tikzpicture}[#1]

\tikzstyle{nodestyle} = [doublecirclestyle];

\begin{scope} [every node/.style=doublecirclestyle]
	\node[nodestyle,colorAnsubgraph] (A) at (0.3,1) {1};
	\node (Adown) at (0.3,-1) {4};
	\node[nodestyle,colorAnsubgraph] (B) at (2,0) {2};
	\node[nodestyle,colorAnsubgraph] (C) at (4,0) {3};
	\node[nodestyle,colorAnsubgraph] (D) at (6,0) {};
	\node[nodestyle,colorAnsubgraph] (E) at (8,0) {$n$};
	\ifthenelse{#3=0 \AND #4=0 \AND #5=0 \AND #6=0}{}{	\node (new) at (2,2) {#2};}
\end{scope}

\node (labelFixed) at (0,0) {$\omega$};

\begin{scope}[every node/.style={fill=white}, every edge/.style={draw,very thick}]
	\path [-,colorAnsubgraph] (A) edge (B);
	\path [-] (Adown) edge (B);
	\path [-,colorAnsubgraph] (B) edge (C);
	\path [dashed,colorAnsubgraph] (C) edge (D);
	\path [-,colorAnsubgraph] (D) edge (E);
\end{scope}

\begin{scope}[every node/.style={fill=white}, every edge/.style={draw,very thick}]
	\arrowedge[#3](new,A)(bend right,newarrowstyle1)
	\arrowedge[#4](new,B)(newarrowstyle1)
	\arrowedge[#5](new,C)(bend left,newarrowstyle1)
	\arrowedge[#6](new,Adown)(bend right=10,newarrowstyle1)
\end{scope}
\end{tikzpicture}
}

\def\sdiagramJ[#1](#2)(#3,#4,#5,#6) { % \diagramJ[draw options](edge new-1, edge new-2, edge new-3)
\vcenter{\hbox{\begin{tikzpicture}[#1]

\tikzstyle{nodestyle} = [doublecirclestyle];

\begin{scope} [every node/.style=doublecirclestyle]
	\node[nodestyle,colorAnsubgraph] (A) at (0.3,1) {1};
	\node (Adown) at (0.3,-1) {4};
	\node[nodestyle,colorAnsubgraph] (B) at (2,0) {2};
	\node[nodestyle,colorAnsubgraph] (C) at (4,0) {3};
	\node[nodestyle,colorAnsubgraph] (D) at (6,0) {};
	\node[nodestyle,colorAnsubgraph] (E) at (8,0) {$n$};
	\IfEq{#2}{}{}{\node (new) at (2,2) {#2};}
\end{scope}

\node (labelFixed) at (0,0) {$\omega$};

\begin{scope}[every node/.style={fill=white}, every edge/.style={draw,very thick}]
	\path [-,colorAnsubgraph] (A) edge (B);
	\path [-] (Adown) edge (B);
	\path [->] (A) edge (Adown);
	\path [-,colorAnsubgraph] (B) edge (C);
	\path [dashed,colorAnsubgraph] (C) edge (D);
	\path [-,colorAnsubgraph] (D) edge (E);
\end{scope}

\IfEq{#2}{}{}{
	\begin{scope}[every node/.style={fill=white}, every edge/.style={draw,very thick}]
		\arrowedge[#3](new,A)(bend right,	newarrowstyle1)
		\arrowedge[#4](new,B)(newarrowstyle1)
		\arrowedge[#5](new,C)(bend left,newarrowstyle1)
		\arrowedge[#6](new,Adown)(bend right=10,newarrowstyle1)
	\end{scope}
}
\end{tikzpicture}}}
}

\def\sdiagramJthree[#1](#2)(#3,#4,#5)(#6,#7,#8) { % \diagramJ[draw options](edge new-1, edge new-2, edge new-3)
\vcenter{\hbox{\begin{tikzpicture}[#1]

\tikzstyle{nodestyle} = [doublecirclestyle];

\begin{scope} [every node/.style=doublecirclestyle]
	\node[nodestyle,colorAnsubgraph] (A) at (0.3,1) {1};
	\node (Adown) at (0.3,-1) {3};
	\node[nodestyle,colorAnsubgraph] (B) at (2,0) {2};
	\IfEq{#2}{}{}{\node (new) at (2,2) {#2};}
\end{scope}

\begin{scope}[every node/.style={fill=white}, every edge/.style={draw,very thick}]
	\arrowedge[#6](A,B)(colorAnsubgraph)
	\arrowedge[#7](Adown,B)()
	\arrowedge[#8](A,Adown)()
\end{scope}

\IfEq{#2}{}{}{
	\begin{scope}[every node/.style={fill=white}, every edge/.style={draw,very thick}]
		\arrowedge[#3](new,A)(bend right,	newarrowstyle1)
		\arrowedge[#4](new,B)(newarrowstyle1)
		\arrowedge[#5](new,Adown)(bend right=10,	newarrowstyle1)
	\end{scope}
}
\end{tikzpicture}}}
}

\def\sdiagramAloop[#1](#2)(#3,#4,#5,#6,#7,#8){
\vcenter{\hbox{\begin{tikzpicture}[#1]

	\node (left) at (-4,0) {};
	\node (right) at (4,0) {};
	
\begin{scope} [every node/.style=doublecirclestyle] 
	\node[double,circle,thick] (jp1) at (-1,0) {};
	\node[label=below:$j_{i}$,double,circle,thick]  (j) at (-2,0) {};
	\node[double,circle,thick]  (jm1) at (-3,0) {};
	\node[double,circle,thick]  (kp1) at (3,0) {};
	\node[label=below:$j_{i+1}$,double,circle,thick]  (k) at (2,0) {};
	\node[double,circle,thick]  (km1) at (1,0) {};
	\node[double,circle,thick] (new) at (0,2) {#2};
\end{scope}

\begin{scope}[every node/.style={fill=white}, every edge/.style={draw,very thick}]
	\path [-] (jm1) edge (j);
	\path [-] (j) edge (jp1);
	\path [dashed] (jp1) edge (km1);
	\path [-] (km1) edge (k);
	\path [-] (k) edge (kp1);
	\path [dashed] (left) edge (jm1);
	\path [dashed] (kp1) edge (right);
\end{scope}

\begin{scope}[every node/.style={fill=white}, every edge/.style={draw,very thick}]
	\arrowedge[#3](new,jm1)(bend right)
	\arrowedge[#4](new,j)(bend right=20)
	\arrowedge[#5](new,jp1)()
	\arrowedge[#6](new,km1)()
	\arrowedge[#7](new,k)(bend left=20)
	\arrowedge[#8](new,kp1)(bend left)
\end{scope}

\end{tikzpicture}}}
}

\def\sdiagramAleft[#1](#2)(#3,#4,#5,#6,#7,#8){
\vcenter{\hbox{\begin{tikzpicture}[#1]

	\node (right) at (4,0) {};
	
\begin{scope} [every node/.style=doublecirclestyle] 
	\node[double,circle,thick] (jp1) at (-1,0) {};
	\node[double,circle,thick]  (j) at (-2,0) {};
	\node[label=below:1,double,circle,thick]  (jm1) at (-3,0) {};
	\node[double,circle,thick]  (kp1) at (3,0) {};
	\node[label=below:$j_{i+1}$,double,circle,thick]  (k) at (2,0) {};
	\node[double,circle,thick]  (km1) at (1,0) {};
	\node[double,circle,thick] (new) at (0,2) {#2};
\end{scope}

\begin{scope}[every node/.style={fill=white}, every edge/.style={draw,very thick}]
	\path [-] (jm1) edge (j);
	\path [-] (j) edge (jp1);
	\path [dashed] (jp1) edge (km1);
	\path [-] (km1) edge (k);
	\path [-] (k) edge (kp1);
	\path [dashed] (kp1) edge (right);
\end{scope}

\begin{scope}[every node/.style={fill=white}, every edge/.style={draw,very thick}]
	\arrowedge[#3](new,jm1)(bend right)
	\arrowedge[#4](new,j)(bend right=20)
	\arrowedge[#5](new,jp1)()
	\arrowedge[#6](new,km1)()
	\arrowedge[#7](new,k)(bend left=20)
	\arrowedge[#8](new,kp1)(bend left)
\end{scope}

\end{tikzpicture}}}
}

\def\sdiagramAreduced[#1](#2)(#3,#4)(#5,#6){ %\sdiagramAreduced[options] (newnode) (edge1, edge2) (node1, node2)
\vcenter{\hbox{\begin{tikzpicture}[#1]
	
\begin{scope} [every node/.style=doublecirclestyle] 
	\IfEq{#5}{1}{}
		{\node[label=below:1,		double,circle,thick]  (jm1) at (-3,0) {};}
	\node[label=below:$#5$,	double,circle,thick] (jp1) at (-1,0) {};
	\node[label=below:$#6$,	double,circle,thick]  (km1) at (1,0) {};
	\node[label=below:$n$,	double,circle,thick]  (kp1) at (3,0) {};
	\node[double,circle,thick] (new) at (0,2) {#2};
\end{scope}

\begin{scope}[every node/.style={fill=white}, every edge/.style={draw,very thick}]
	\IfEq{#5}{1}{}{\path[dashed] (jm1) edge (jp1);}
	\IfEq{#5-#6}{-1}{\path [-] (jp1) edge (km1);}
		{\path [dashed] (jp1) edge (km1);}
	\path [dashed] (jp1) edge (km1);
	\path [dashed] (km1) edge (kp1);
\end{scope}

\begin{scope}[every node/.style={fill=white}, every edge/.style={draw,very thick}]
	\arrowedge[#3](new,jp1)()
	\arrowedge[#4](new,km1)()
\end{scope}

\end{tikzpicture}}}
}

\def\sdiagramAreducedshort[#1](#2)(#3,#4)(#5,#6){ %\sdiagramAreduced[options] (newnode) (edge1, edge2) (node1, node2)
\vcenter{\hbox{\begin{tikzpicture}[#1]
	
\begin{scope} [every node/.style=doublecirclestyle] 
	\IfEq{#5}{1}{}
		{\node[label=below:1,		double,circle,thick]  (jm1) at (-3,0) {};}
	\node[label=below:$#5$,	double,circle,thick] (jp1) at (-1,0) {};
	\node[label=below:$#6$,	double,circle,thick]  (km1) at (1,0) {};
	\node[label=below:$n$,	double,circle,thick]  (kp1) at (3,0) {};
	\node[circle,thick] (new) at (0,2) {#2};
\end{scope}

\begin{scope}[every node/.style={fill=white}, every edge/.style={draw,very thick}]
	\IfEq{#5}{1}{}{\path[dashed] (jm1) edge (jp1);}
	\IfEq{#5-#6}{-1}{\path [-] (jp1) edge (km1);}
		{\path [dashed] (jp1) edge (km1);}
	\path [dashed] (jp1) edge (km1);
	\path [dashed] (km1) edge (kp1);
\end{scope}

\begin{scope}[every node/.style={fill=white}, every edge/.style={draw,very thick}]
	\arrowedge[#3](new,jp1)()
	\arrowedge[#4](new,km1)()
\end{scope}

\end{tikzpicture}}}
}

\def\sdiagramLoop[#1](#2)(#3,#4)(#5,#6){ %\sdiagramLoop[options] (newnode) (edge1, edge2) (node1, node2)
\vcenter{\hbox{\begin{tikzpicture}[#1]

\begin{scope} [every node/.style=doublecirclestyle] 
	\node[label=below:$#5$,	double,circle,thick] (left) at (-2,0) {};
	\node[label=below:$#6$,	double,circle,thick]  (right) at (2,0) {};
	\node[double,circle,thick] (new) at (0,2) {#2};
\end{scope}

\begin{scope}[every node/.style={fill=white}, every edge/.style={draw,very thick}]
	\path [dashed] (left) edge (right);
\end{scope}

\begin{scope}[every node/.style={fill=white}, every edge/.style={draw,very thick}]
	\arrowedge[#3](new,left)()
	\arrowedge[#4](new,right)()
\end{scope}

\end{tikzpicture}}}
}

\def\sdiagramDoubleLoop[#1](#2)(#3,#4,#5)(#6,#7,#8){ %\sdiagramDoubleLoop[options] (newnode) (edge1, edge2,edge3) (node1, node2, node3)
\vcenter{\hbox{\begin{tikzpicture}[#1]

\begin{scope} [every node/.style=doublecirclestyle] 
	\node[label=below:$#6$,	double,circle,thick] (left) at (-2,0) {};
	\node[label=below:$#7$,	double,circle,thick]  (center) at (0,0) {};
	\node[label=below:$#8$,	double,circle,thick]  (right) at (2,0) {};
	\node[double,circle,thick] (new) at (0,2) {#2};
\end{scope}

\begin{scope}[every node/.style={fill=white}, every edge/.style={draw,very thick}]
	\IfEq{#7}{2}{\path [-] (left) edge (center);}
		{\path [dashed] (left) edge (center);}
	\IfEq{#7}{r-2}{\path [-] (center) edge (right);}
		{\path [dashed] (center) edge (right);}
\end{scope}

\begin{scope}[every node/.style={fill=white}, every edge/.style={draw,very thick}]
	\arrowedge[#3](new,left)()
	\arrowedge[#4](new,center)()
	\arrowedge[#5](new,right)()
\end{scope}

\end{tikzpicture}}}
}

\title{
\begin{center}
The ABCDEFGJK of maximally strongly coupled $\mathcal{N}=2$ SCFTs
\end{center}
}

\author[a]{Antonio %AMARITO
Amariti}
\author[b,c]{and Simone %ALESSIO
 Rota}
\affiliation[a]{INFN, Sezione di Milano, Via Celoria 16, I-20133 Milano, Italy}

\affiliation[b]{SISSA, Via Bonomea 265, 34136 Trieste, Italy}

\affiliation[c]{INFN, Sezione di Trieste, Via Valerio 2, 34127 Trieste, Italy}

\emailAdd{antonio.amariti@mi.infn.it,  srota@sissa.it}

\abstract{
We classify all possible charge lattices and 1-form symmetry groups for $\mathcal{N}=2$ SCFTs with characteristic dimension $\varkappa \neq \{1,2\}$. 
For rank-$r$ SCFTs that are not stacks of lower rank theories the order of the 1-form symmetry group can be 1,2,3,4 and $r+1$.
As an application of the classification we show that $\mathcal{N}=2$ $S$-folds and Minahan-Nemeschansky SCFTs have trivial 1-form symmetry.
We find two sporadic lattices compatible with the Coulomb branch geometries $\mathbb{C}^5/G_{33}$ and $\mathbb{C}^6/G_{34}$ that are not realized by any known SCFT. The former, if realized, would have a $\mathbb{Z}_2$ 1-form symmetry and a non-invertible topological defect.
}

\begin{document}

\maketitle

%%%%%%%%%%%%%%%%%%%%%%%%%%%%%%%%%%%%%%%%%%%%%%%%%%%%%%%%%%%%%%%%
\section{Introduction}

% introduction and motivation
In the landscape of quantum field theories a special role is played by conformal field theories (CFTs), which are fixed points under the RG flow. The conformal symmetry imposes general constraints which render CFTs a favorable object for theoretical analysis. 
%{\color{red} TOGLIERE? -> It is desirable to exploit the conformal symmetry to study general features of CFTs.
%When a CFT enjoys supersymmetry the amount of constraints that a theory must satisfy is significantly larger}
Supersymmetry further constrains a CFT and, with enough supercharges, a full classification of SCFTs may become viable. For example in $4d$ it is believed that SCFTs with 16 supercharges ($\mathcal{N}=4$) are fully classified by maximally supersymmetric SYM theories. Lowering the amount of supersymmetry general features of SCFTs becomes increasingly challenging to understand and in the case of minimal supersymmetry, $\NN=1$, a classification appears to be out of reach with the available technology.

% N=2 - global structure from the infrared
A bottom up analysis of $\NN=2$ SCFTs, though far from providing a full classification of these theories, has produced promising results \cite{Argyres:2015gha,Argyres:2016xmc,Argyres:2016xua,Argyres:2015ffa,Martone:2021ixp,Kaidi:2021tgr,Xie:2015rpa,Chen:2016bzh,Wang:2016yha,Chen:2017wkw,Xie:2021hxd,Cecotti:2021ouq,Martone:2020nsy,Argyres:2020wmq,Martone:2021drm,Argyres:2020nrr,Caorsi:2018zsq,Argyres:2018zay,Caorsi:2018ahl,Caorsi:2019vex,Argyres:2019yyb,Argyres:2018wxu,Bourget:2018ond,Argyres:2019ngz,Argyres:2017tmj,Nishinaka:2016hbw,Argyres:2023eij,Closset:2023pmc,Furrer:2024zzu}. 
Most notably a full classification\footnote{The rank-1 classification is valid under the assumption that no rank-0 interacting theory exists. The existence of interacting $\NN=2$ SCFTs without a Coulomb Branch has not been ruled out completely, although throughout this paper we will assume that no such theory exists.} of $\NN=2$ rank-1 theories has been developed \cite{Argyres:2016xmc, Argyres:2015ffa, Argyres:2015gha, Argyres:2016xua} and much progress has been achieved for the rank-2 case (see \cite{Argyres:2022lah,Argyres:2022puv,Argyres:2022fwy,Argyres:2018zay} and references therein). 

The case of  $4d$ $\mathcal{N}=3$ SCFTs deserves a further comment. 
SCFTs with 12 supercharges that are not $\NN=4$ were first discovered in \cite{Garcia-Etxebarria:2015wns} as the low energy theory of a stack of D3-branes probing an $S$-fold singularity in Type IIB. This construction was later generalized producing new $\NN=3$ \cite{Garcia-Etxebarria:2016erx},  $\NN=2$ \cite{Apruzzi:2020pmv, Giacomelli:2020jel, Giacomelli:2020gee, Heckman:2020svr,Giacomelli:2024dbd} and $\NN=1$ \cite{Giacomelli:2023qyc} SCFT. As pointed out in \cite{Aharony:2016kai}, the moduli space of all known $\mathcal{N}=3$ SCFTs, modulo discrete gauging, is given by an orbifold of $\mathbb{C}^r$ by a crystallographic complex reflection group (CCRG), suggesting a correspondence between $\NN=3$ SCFTs and CCRGs\footnote{See also \cite{Tachikawa:2019dvq,Deb:2024zay} for analogous considerations in various dimensions.}. 
One is then lead to consider putative CB geometries given by orbifolds $\mathbb{C}^r/G$, with $G$ one of the CCRGs classified by Shephard and Todd, see \cite{CCRG} for a comprehensive review.
SCFTs realizing some of these CBs where lated discovered in \cite{Kaidi:2022lyo}, and the correspondence was refined in \cite{Amariti:2023qdq}. 
Nevertheless there are no known SCFTs realizing the CB geometries $\mathbb{C}^r/G$ with $G= G_{24},G_{29},G_{33}$ and $G_{34}$, therefore the correspondence between CCRGs and $\mathcal{N}=3$ SCFTs remains an open question.

% N=2 - global structure from the infrared
One of the key aspects underlying these classification efforts is the presence of the Coulomb branch (CB), a moduli space of vacua where the $U(1)_R$ subgroup of the $\NN=2$ R-symmetry group is spontaneously broken while the $SU(2)_R$ subgroup is preserved. On a generic point of the \CB  the low energy physics is described by a $U(1)^r$ gauge theory coupled to massive electrically and magnetically charged particles. $r$ is the rank of the SCFT and is equal to the complex dimension of the CB.
The masses of charged particles is bounded from below by a multiple of the absolute value of their central charge $Z(e,m,\mathbf{u})$, with BPS states saturating the bound. As one varies the position $\mathbf{u}$ and the CB, BPS states will becomes massless whenever their central charge vanishes. At these special points the free $U(1)^r$ description breaks down and the CB develops a singularity where non-trivial interacting dynamics take place.
%Powerful techniques have been developed to study the spectrum of charged BPS states in $\NN=2$ theories, 
%{\color{red}for example Seiberg-Witten solutions \cite{} and BPS quivers \cite{} (qui si rischia di raddoppiare le citazioni, magari toglierlo)}, 
%still it remains a challenging task to determine the full BPS spectrum. 
Analyzing the spectrum of BPS operators is a challenging task, and
a simpler object to study in this regard is the charge lattice $\Gamma$, the set of electromagnetic charges of all charged states. 
The charge lattice is equipped with a Dirac pairing $J$, an integer-valued bilinear antisymmetric map:
\begin{equation}
J: \Gamma \times \Gamma \to \mathbb{Z}
\end{equation}
It was shown in \cite{DelZotto:2022ras,Argyres:2022kon} that $\Gamma$ and $J$ are closely related to the 1-form symmetry \cite{Gaiotto:2010be,Gaiotto:2014kfa,Aharony:2013hda} group $G^{(1)}$ of the theory. 
The order of $G^{(1)}$ is given by:
\begin{equation}
|G^{(1)}| = |\Pf(J)|
\end{equation}
%where with an abuse of notation we denote as $J$ the matrix representing the Dirac pairing on the basis $\gamma_i$.
This relation was exploited in \cite{DelZotto:2022ras} to study the global forms of $\NN=2$ theories with a BPS quiver and later in \cite{Amariti:2023hev} to study $S$-fold theories. 
See also \cite{Garding:2023unh} for an application of these ideas to class-S theories.

% x != 1,2 and graphs
In this paper we show that
a general analysis of the charge lattice becomes viable for SCFTs with characteristic dimension $\varkappa\neq \{1,2\}$ \cite{Cecotti:2021ouq}. The characteristic dimension is an invariant of $\NN=2$ SCFTs taking 8 possible values.
\begin{equation}
\varkappa \in \left\{1,2,3,\frac{2}{3},4,\frac{3}{4}, 6, \frac{5}{6} \right\}
\end{equation}
When $\varkappa\neq \{1,2\}$ the $U(1)_R$ symmetry is broken to a $\zz_k$ discrete subgroup on the CB, with:
\begin{equation}
k=  \left\{
\renewcommand\arraystretch{1.2}
\begin{array}{ll}
3 &\qquad \varkappa = 3,\frac{2}{3}
\\
4 & \qquad\varkappa=4,\frac{3}{4}
\\
6 & \qquad\varkappa = 6, \frac{5}{6}
\end{array}
\right.
\renewcommand\arraystretch{1}
\end{equation}

% $k=4$ for $\varkappa=4,\frac{3}{4}$, $k=3$ for $\varkappa = 3,\frac{2}{3}$ and $k=6$ for $\varkappa = 6, \frac{5}{6}$. 
%In the case of $\varkappa=2,\frac{2}{3}$ the $\zz_3$ symmetry combines with CPT into a $\zz_6$ group.
This symmetry acts non-trivially on the electromagnetic charges, mapping a state with central charge $Z[q]$ to a mutually non-local state with central charge $\zeta_k Z[q]$ where $\zeta_k=e^{2\pi i /k}$. 
%These two states are mutually non-local therefore 
When a charged state becomes massless then also another state, mutually nonlocal with respect to the first one, becomes massless as well. On the CB this implies that the dynamics supported on the singularities is always described by a strongly coupled SCFT. 
Furthermore the charge lattice, which is a real lattice in $\mathbb{R}^{2r}$, can be identified with a complex lattice in $\mathbb{C}^r$. On the complex lattice the $\zz_k$ symmetry acts with an overall $\zeta_k$ factor while the Dirac pairing can be written as\footnote{Here and in the rest of the paper with an abuse of notation we denote the charges in the real charge lattice and their images in $\mathbb{C}^r$ with the same symbol.}:
\begin{equation}	\label{eq:Hdef_formula}
\langle q,p \rangle = \frac{1}{\zeta_k - \overline{\zeta_k}} (H(q,p)- H(p,q))
\end{equation}
where $H$ is a non-degenerate positive definite Hermitian form valued in $\zz[\zeta_k]$. 

The fact that codimension-1 singularities support SCFTs combined with the classification of rank-1 SCFTs strongly constrains the charge lattice. 
%It is useful to choose a basis for the charge lattice and represent $H$ as an Hermitian matrix. 
To see this let us choose a basis given by charges $q_i$ and $\zeta_k q_i$ that become massless on some codimension-1 singularity labeled by $i$. 
%{\color{red}
For simplicity in this paper we assume that the spectrum of charged BPS states provides a basis for the charge lattice.
In more general contexts there are known theories whose charge lattice is not spanned by the BPS spectrum, see \cite{DelZotto:2022ras} for a discussion on this topic. 
However, if an SCFT had additional charges not spanned by the BPS states then the charge lattice would be a refinement of one of the lattices discussed here\footnote{Refinements of the charge lattices of type $A_n$, $BC_n$, $D_n$ and $E_n$ can be obtained with techniques analogous to the ones used to determine the global forms of $\NN=4$ SYM with the corresponding gauge algebra \cite{Aharony:2013hda}. The refinements of the $J_n$ lattices have been analyzed in \cite{Amariti:2023hev} in the context of $S$-fold SCFTs and the refinements of $K_5$ are analyzed in Section \ref{sec:K5}. The other charge lattices do not admit refinements. See also \cite{Argyres:2022kon} for a general algorithm to determine maximal refinements.}.
%}
The charges becoming massless at a singularity generate the charge lattice of the rank-1 SCFTs supported on that singularity, therefore we can relate the 1-form symmetry group $G^{(1)}_{\text{rank-1}}$ of the rank-1 SCFT with the diagonal elements of $H$:
\begin{equation}
|G^{(1)}_{\text{rank-1}}|=\Pf(J^{(1)}) = \langle q, \zeta_k q \rangle = H(q,q) = 1,2
\end{equation}
where we used the fact that the 1-form symmetry of a rank-1 $\NN=2$ SCFT is either $\zz_2$ in the case of $\NN=4$ $SU(2)$ SYM or trivial otherwise \cite{Argyres:2022kon}.
In this basis diagonal elements of $H$ can be either 1 or 2 and the off-diagonal elements are bounded by the Cauchy-Schwarz inequality and can only take a finite set of values, listed in Table \ref{tab:possibleHij}.

All in all, at any finite rank there is a finite set of possible Dirac pairings that can be realized by $\NN=2$ SCFTs with $\varkappa \neq\{1,2\}$. Physically, this is due to the fact that the classification of rank-1 SCFTs, imposing $H_{ii}=1,2$, and the Dirac quantization condition, imposing $H_{ij}\in \zz[\zeta_k]$, strongly constrain the charge lattice.
In this paper we fully classify the inequivalent Hermitian matrices $H$ satisfying the conditions above, which in turn provides a full list of possible charge lattices for $\NN=2$ SCFTs with $\varkappa \neq\{1,2\}$. 
The results for SCFTs that are not stacks of lower rank SCFTs, which we denote as irreducible SCFTs in this paper, are summarized in a diagram notation in Table \ref{tab:results}.
There we give examples of SCFTs that realize the various lattices, if any is known. 
We reproduce the charge lattices of fluxless $S_{3,3}$-fold SCFTs \cite{Imamura:2016udl,Agarwal:2016rvx,Amariti:2023hev}, denoted as $J_n$ and fluxfull $S$-fold theories, denoted as $\overline{A_n}$.
It should be noted that the characterization of the charge lattices given above\footnote{Equivalently, the set of Hermitian matrices $H$ that we consider.} only relies on the presence of a $\zz_k$ symmetry of the charge lattice, with $k=3,4,6$, acting on the central charges of charged stated as multiplication by $\zeta_k$. 
The charge lattice of
$\NN=4$ SYM enjoys such a symmetry at specific values of the complexified gauge coupling  that are fixed by subgroups of the S-duality group.
Notice that in the case of non-simply-laced algebras this symmetry of the charge lattice is not necessarily a symmetry of the full theory. 
But, crucially, this fact does not affect the discussion above and in our classification we find the charge lattices of
%Accordingly our classification also includes the charge lattices of 
all $\NN=4$ SYM theories that we denote with the name of the corresponding gauge algebra\footnote{The charge lattice of $G_2$ is given by $\overline{A_2}$. This is compatible with the fact that $G_2$ $\NN=4$ SYM is engineered by two D3-branes probing an $S_6$-fold \cite{Aharony:2016kai}, therefore it appears as the first element of an infinite family of SCFTs. }. Compatibly with S-duality, $\NN=4$ SYM with gauge algebras $\mathfrak{b}_n$ and $\mathfrak{c}_n$ have the same charge lattice that we denote as $BC_n$.
We find two sporadic lattices denoted as $K_5$ and $K_6$. To the best of our knowledge there is no known SCFT realizing $K_5$ and $K_6$, though they are compatible with putative CB geometries introduced in \cite{Kaidi:2022lyo}. 

For all charge lattices in the classification we compute the possible 1-form symmetry group $G^{(1)}$ whose order is given by $\det(H)$. We find that the possible values for the order of the 1-form symmetry group of an irreducible rank-$r$ SCFT with $\varkappa \neq\{1,2\}$ are:
\begin{equation}
|G^{(1)}_{\text{rank-}r}|=1,2,3,4,r+1
\end{equation}
Furthermore for rank-2 theories the only allowed values are $|G^{(1)}_{\text{rank-}2}|=1,2,3$.

We show that if an irreducible SCFTs with $\varkappa\neq\{1,2\}$ has a singularity supporting any SCFT other than $SU(2)$ $\NN=4$ then it the charge lattice is $\overline{A_n}$ and the 1-form symmetry is trivial. This implies that higher-rank Minahan-Nemeschansky theories and most $\NN=2$ $S$-folds have trivial 1-form symmetry. 
 $K_5$ and $K_6$ represent new sporadic charge lattices and it is interesting to study some of the properties of a putative SCFTs that would realize them.
The $K_6$ has trivial 1-form symmetry while the $K_5$ has a $\zz_2$ 1-form symmetry. 
We show that the $K_5$ lattice has 3 global forms corresponding to the maximal sets of line operators compatible with the Dirac quantization condition. Furthermore we show that the $\zz_3$ global symmetry exchanges the global forms and should be combined with a half-space gauging interface to produce a proper topological defect. 
This topological defect is non-invertible and is analogous to the triality defect of $SU(2)$ $\NN=4$ SYM
%\cite{Kaidi:2021xfk,Bhardwaj:2022yxj,Choi:2022zal,Kaidi:2022uux} 
\cite{Choi:2021kmx,Kaidi:2021xfk,Bhardwaj:2022yxj,Choi:2022zal, Kaidi:2022uux,Bhardwaj:2022lsg,Bartsch:2022mpm,Antinucci:2022vyk,Bhardwaj:2022kot,Bartsch:2022ytj,Bhardwaj:2022maz,Heckman:2022xgu,Nardoni:2024sos,DelZotto:2024tae}
(see also \cite{Bhardwaj:2023kri} and references therein).

This paper is organized as follows. In Section \ref{sec:2} we introduce the graphical notation used throughout the paper and outline the strategy used for the classification of charge lattices for $\NN=2$ SCFTs with $\varkappa\neq\{1,2\}$. 
The strategy proceeds “by induction” on the rank of the SCFT and we apply it to classify the lattices for rank-2 and rank-3 SCFTs. 
The generalization to all-rank is carried out in Section \ref{sec:3}. The results of the classification are collected in Table \ref{tab:results}.
In Section \ref{sec:4} we exploit 
these results to analyze the generalized symmetries of  $\NN=2$ SCFTs with $\varkappa\neq\{1,2\}$, focussing on $\NN=2$ $S$-folds and Minahan-Nemeschansky theories.
Section \ref{sec:conclusions} contains some closing remarks and possible future directions.

%%%%%%%%%%%%%%%%%%%%%%%%%%%%%%%%%%%%%%%%%%%%%%%%%%%%%%%%%%%%%%%%
%%%%%%%%%%%%%%%%%%%.         RESULTS.          %%%%%%%%%%%%%%%%%%%%%
\begin{table}
\centering{
$H_{ii} = 2: \quad \vcenter{\hbox{\tikz{\node[double,circle,thick,draw] (A) at (0,0) {};}}}
\qquad\qquad H_{ii} = 1: \quad \vcenter{\hbox{\tikz{\node[circle,thick,draw] (A) at (0,0) {};}}}
\qquad\qquad H_{ij} = \alpha: \quad  \vcenter{\hbox{\tikz{\node[dashed,gray,circle,thick,draw,minimum size = 10pt] (A) at (0,0) {i};\node[dashed,gray,circle,thick,draw,minimum size = 10pt] (B) at (2,0) {j};\arrowedge[\alpha](A,B)(every node/.style={fill=white}, very thick) }}}
 $
 }
\renewcommand{\arraystretch}{1.8}
\def\scalegraphstable{0.7}
\begin{equation*}
   \makebox[\displaywidth]{$\displaystyle
\begin{array}{|cccc|c|}
\hline
&\text{diagram} & \makebox[3cm]{rank} & \makebox[2cm]{$ |G^{(1)}| $}& \text{theory example}
\\ \hline
\multicolumn{5}{|c|}{\varkappa \in \left\{3,\frac{2}{3},4,\frac{3}{4}, 6, \frac{5}{6} \right\} }
\\ \hline
A_n: &
\vcenter{\hbox{\begin{tikzpicture}[scale=\scalegraphstable]
%nodes
\begin{scope} [every node/.style={double,circle,thick,draw}]
\node (A) at (0,0) {};
\node (C) at (2,0) {};
\node (D) at (4,0) {};
\node (E) at (6,0) {};
\node (F) at (8,0) {};
\end{scope}
%edges
\begin{scope}[every node/.style={fill=white,circle}, every edge/.style={draw,very thick}]
\path [-] (A) edge (C);
\path [-] (C) edge (D);
\path [-] (E) edge (F);
\path [dashed] (D) edge (E);
\end{scope}
\end{tikzpicture}}}
& n & n+1 & \text{$\mathfrak{a}_n$ $\mathcal{N}=4$ SYM}
\\
D_n: &
\vcenter{\hbox{\begin{tikzpicture}[scale=\scalegraphstable]
%nodes
\begin{scope} [every node/.style={double,circle,thick,draw}]
\node (A) at (0,1.15) {};
\node (B) at (0,-1.15) {};
\node (C) at (2,0) {};
\node (D) at (4,0) {};
\node (E) at (6,0) {};
\node (F) at (8,0) {};
\end{scope}
%edges
\begin{scope}[every node/.style={fill=white,circle}, every edge/.style={draw,very thick}]
\path [-] (A) edge (C);
\path [-] (B) edge (C);
\path [-] (C) edge (D);
\path [-] (E) edge (F);
\path [dashed] (D) edge (E);
\end{scope}
\end{tikzpicture}}}
& n\geq3&4 &\text{$\mathfrak{d}_n$ $\mathcal{N}=4$ SYM}
\\
E_n: &
\vcenter{\hbox{\begin{tikzpicture}[scale=\scalegraphstable]
%nodes
\begin{scope} [every node/.style={double,circle,thick,draw}]
\node (A) at (0,0) {};
\node (C) at (2,0) {};
\node (D) at (4,0) {};
\node (E) at (6,0) {};
\node (F) at (8,0) {};
\node (B) at (4,2) {};
\end{scope}
%edges
\begin{scope}[every node/.style={fill=white,circle}, every edge/.style={draw,very thick}]
\path [-] (A) edge (C);
\path [-] (C) edge (D);
\path [-] (E) edge (F);
\path [-] (B) edge (D);
\path [dashed] (D) edge (E);
\end{scope}
\end{tikzpicture}}}
& n=6,7,8 & 9-n &\text{$\mathfrak{e}_n$ $\mathcal{N}=4$ SYM}
\\
\overline{A_n}: &
\vcenter{\hbox{\begin{tikzpicture}[scale=\scalegraphstable]
%nodes
\node[circle,thick,draw] (A) at (0,0) {};
\begin{scope} [every node/.style={double,circle,thick,draw}]
\node (C) at (2,0) {};
\node (D) at (4,0) {};
\node (E) at (6,0) {};
\node (F) at (8,0) {};
\end{scope}
%edges
\begin{scope}[every node/.style={fill=white,circle}, every edge/.style={draw,very thick}]
\path [-] (A) edge (C);
\path [-] (C) edge (D);
\path [-] (E) edge (F);
\path [dashed] (D) edge (E);
\end{scope}
\end{tikzpicture}}}
& n & 1 & \begin{gathered}\text{fluxful $S$-folds \cite{Aharony:2016kai},}\\ \text{ MN theories} \end{gathered}
\\ \hline
\multicolumn{5}{|c|}{\varkappa \in \left\{3,\frac{2}{3}, 6, \frac{5}{6} \right\} }
\\ \hline 
J_n: &
\vcenter{\hbox{\begin{tikzpicture}[scale=\scalegraphstable]
%nodes
\begin{scope} [every node/.style={double,circle,thick,draw}]
\node (A) at (0,1.15) {};
\node (B) at (0,-1.15) {};
\node (C) at (2,0) {};
\node (D) at (4,0) {};
\node (E) at (6,0) {};
\node (F) at (8,0) {};
\end{scope}
%edges
\begin{scope}[every node/.style={fill=white}, every edge/.style={draw,very thick}]
\path [-] (A) edge (C);
\path [-] (B) edge (C);
\path [-] (C) edge (D);
\path [-] (E) edge (F);
\path [->] (A) edge node {$\omega$} (B);
\path [dashed] (D) edge (E);
\end{scope}
\end{tikzpicture}}}
& n\geq3&3 & \text{$S_{3,3}$-folds \cite{Garcia-Etxebarria:2015wns,Aharony:2016kai}}
\\
K_5: &
\diagramKfivesmall[scale=\scalegraphstable,every node/.style=smalldoublecirclestyle](,,,,)
& 5&2 & \text{??}
\\
K_6: &
\diagramKsixsmall[scale=\scalegraphstable](,,,,,)
& 6 & 1 & \text{??}
\\ \hline
\multicolumn{5}{|c|}{\varkappa \in \left\{4, \frac{3}{4} \right\} }
\\ \hline
BC_n: &
\vcenter{\hbox{\begin{tikzpicture}[scale=\scalegraphstable]
%nodes
\begin{scope} [every node/.style={double,circle,thick,draw}]
\node (A) at (0,0) {};
\node (C) at (2,0) {};
\node (D) at (4,0) {};
\node (E) at (6,0) {};
\node (F) at (8,0) {};
\end{scope}
%edges
\begin{scope}[every node/.style={fill=white,circle}, every edge/.style={draw,very thick,above}]
\path [->] (A) edge node {$1+i$} (C);
\path [-] (C) edge (D);
\path [-] (E) edge (F);
\path [dashed] (D) edge (E);
\end{scope}
\end{tikzpicture}}}
& n\geq2&2 & \begin{gathered}\text{$\mathfrak{b}_n$ and $\mathfrak{c}_n$ $\mathcal{N}=4$ SYM},\\ \text{$S_{4,4}$-folds \cite{Garcia-Etxebarria:2015wns,Aharony:2016kai}}\end{gathered}
\\
F_4: &
\vcenter{\hbox{\begin{tikzpicture}[scale=\scalegraphstable]
%nodes
\begin{scope} [every node/.style={double,circle,thick,draw}]
\node (A) at (0,0) {};
\node (C) at (2,0) {};
\node (D) at (4,0) {};
\node (E) at (6,0) {};
\end{scope}
%edges
\begin{scope}[every node/.style={fill=white,circle}, every edge/.style={draw,very thick,above}]
\path [->] (C) edge node {$1+i$} (D);
\path [-] (A) edge (C);
\path [-] (D) edge (E);
\end{scope}
\end{tikzpicture}}}
&4& 1 & \begin{gathered}\text{$\mathfrak{f}_4$ $\mathcal{N}=4$ SYM},\\ \text{$G_{31}$ exceptional $S$-fold \cite{Garcia-Etxebarria:2016erx,Kaidi:2022lyo}}\end{gathered}
\\[15pt] \hline
\end{array}
$}
\end{equation*}
\caption{Charge lattices of irreducible $\NN=2$ SCFTs with $\varkappa\neq\{1,2\}$. The graph notation is the one introduced in  Section \ref{sec:2}. For each lattice we report the rank, given by the number of nodes, the order of the 1-form symmetry group $G^{(1)}$ and examples of theories realizing the lattice, if any is known. Here $\omega = e^{\frac{\pi i}{3}  }$.}
\label{tab:results}
\renewcommand{\arraystretch}{1}
\end{table}
%%%%%%%%%%%%%%%%%%%.         RESULTS.          %%%%%%%%%%%%%%%%%%%%%
%%%%%%%%%%%%%%%%%%%%%%%%%%%%%%%%%%%%%%%%%%%%%%%%%%%%%%%%%%%%%%%%

\pagebreak
%%%%%%%%%%%%%%%%%%%%%%%%%%%%%%%%%%%%%%%%%%%%%%%%%%%%%%%%%%%%%%%%
%\section{Classification of charge lattices for $\varkappa \neq \{1,2\}$}
%\section{Graphs, determinants and all that}
\section{General strategy}
\label{sec:2}
As discussed in \cite{Cecotti:2021ouq} and briefly reviewed in the introduction the charge lattices of $\NN=2$ SCFTs with characteristic dimension $\varkappa \neq \{1,2\}$ can be identified with a rank-$r$ complex lattice.
On the complex lattice the Dirac pairing is encoded in a positive-definite Hermitian from $H$ valued in $\zz[\zeta_k]$ through \eqref{eq:Hdef_formula} and the unbroken $\zz_k$ subgroup of the $U(1)_R$ R-symmetry acts by multiplication by $\zeta_k$. In a basis $\{q_i\}$ associated to massless states on codimension-1 singularities, as described above, $H$ is represented by a $\zz[\zeta_k]$-valued matrix where the diagonal elements are either 1 or 2. 
In this Section we outline the classification procedure of such matrices. 
The classification at all ranks is carried out in Section \ref{sec:3}.
%classify the equivalence classes of such matrices up to change of basis of the charge lattice. 

%%%%%%%%%
%\subsection{General strategy}
Before delving into the classification it is useful to introduce a graphical notation for such matrices. 
This notation is analogous to the use of Dinking diagrams, encoding Cartan matrices, or the graph notation of \cite{Cohen1976} in the context of complex root systems.
To a given rank-$r$ matrix $H$ we associate a graph diagram with $r$ nodes constructed as follows:
\begin{itemize}
\item For each basis element $q_i$ we draw a node. If $H_{ii}=1$ we draw the corresponding node as a circle \singlenode, if $H_{ii}=2$ we draw a double circle \doublenode instead. 
If needed we label the node with the index $i$ of the corresponding charge $q_i$.
We call $q_i$ a \textit{long charge} if $H_{ii}=2$, while we call it a \textit{short charge} if $H_{ii}=1$.

\begin{equation}
H_{ii} = 2: \quad \vcenter{\hbox{\tikz{\node[doublecirclestyle] (A) at (0,0) {$i$};}}}, \quad q_i \textit{ long charge}
\end{equation}
\begin{equation}
H_{ii} = 1: \quad \vcenter{\hbox{\tikz{\node[singlecirclestyle] (A) at (0,0) {$i$};}}}, \quad q_i \textit{ short charge}
\end{equation}
We use a dashed gray circle \tikz{\node[graycirclestyle] at (0,0) {};} to denote a generic node.

\item For each pair $1\leq i,j \leq r$ we choose an order, say $[i,j]$, and draw an arrow from the node $i$ to the node $j$ labelled by $H_{ij} = \overline{H_{ji}}$. If $H_{ij}$ is real we draw an unoriented edge instead and if $H_{ij}=-1$ we omit the label. If $H_{ij}=0$ we do not draw anything between the two nodes.
\begin{equation}
\renewcommand{\arraystretch}{2}
\begin{array}{rl}
H_{ij} = \overline{H_{ji}} = \alpha:& \quad  \vcenter{\hbox{\tikz{\node[graycirclestyle] (A) at (0,0) {$i$};\node[graycirclestyle] (B) at (2,0) {$j$};\arrowedge[\alpha](A,B)(every node/.style={fill=white}, very thick) }}}
\\
H_{ij} = 1:& \quad  \vcenter{\hbox{\tikz{\node[graycirclestyle] (A) at (0,0) {$i$};\node[graycirclestyle] (B) at (2,0) {$j$};\arrowedge[1](A,B)(every node/.style={fill=white}, very thick) }}}
\\
H_{ij} = -1:& \quad  \vcenter{\hbox{\tikz{\node[graycirclestyle] (A) at (0,0) {$i$};\node[graycirclestyle] (B) at (2,0) {$j$};\arrowedge[-1](A,B)(every node/.style={fill=white}, very thick) }}}
\end{array}
\renewcommand{\arraystretch}{1}
\end{equation}
\end{itemize}

In this notation a rank-$r$ charge lattice is encoded in a graph with $r$ nodes where the edges encode the non-vanishing Dirac pairing between charges.
The determinant of a graph is the determinant of the corresponding Hermitian matrix which is always integer because it is real and it is in $\zz[\zeta_k]$. 
The requirement that $H$ be positive definite is equivalent to requiring that each subgraph, including the graph itself, has positive determinant. 
As an example consider the rank-3 fluxless $S$-fold SCFT, whose charge lattice was computed in \cite{Amariti:2023hev}. In a particular basis we have:
\begin{equation}
J = \left(
\begin{array}{cccccc}
 0 & 1 & 0 & 2 & 1 & -1 \\
 -1 & 0 & 0 & 0 & 2 & -1 \\
 0 & 0 & 0 & -1 & -1 & 2 \\
 -2 & 0 & 1 & 0 & 1 & 0 \\
 -1 & -2 & 1 & -1 & 0 & 0 \\
 1 & 1 & -2 & 0 & 0 & 0 \\
\end{array}
\right)
\leftrightarrow
H=
\left(
\begin{array}{ccc}
 2 & \omega & -1 \\
 \overline{\omega}& 2 & -1 \\
 -1 & -1 & 2 \\
\end{array}
\right)
\leftrightarrow
\vcenter{\hbox{\begin{tikzpicture}
%nodes
\begin{scope} [every node/.style=doublecirclestyle]
\node (A) at (0,1.15) {$1$};
\node (B) at (0,-1.15) {$2$};
\node (C) at (2,0) {$3$};
\end{scope}
%edges
\begin{scope}[every node/.style={fill=white}, every edge/.style={draw,very thick}]
\path [-] (A) edge (C);
\path [-] (B) edge (C);
\path [->] (A) edge node {$\omega$} (B);
\end{scope}
\end{tikzpicture}}}
\end{equation}

% change of basis and moves on the graphs
The matrices $H$ and consequently the diagrams are base dependent and are not uniquely fixed by a choice of charge lattice. 
Indeed the main computational task carried out in this Section is to determine when two diagrams are related by a change of basis.
After a change of basis the matrix $H$ may have diagonal entries different from 1 or 2, but in this paper we only consider changes of basis that preserve this property. In particular we consider the following basic moves:
\begin{itemize}
\item Shifts in the phase of a charge $q_i$:
\begin{equation}
q_i \to \zeta_k q_i
\end{equation}
 This move corresponds to exchanging the basis element $q_i$ with one of the charges in the $\zz_k$ orbit of $q_i$. 
Schematically the corresponding move on the diagram is:

\begin{equation}	\label{eq:move1}
\vcenter{\hbox{\begin{tikzpicture}[scale=2]
%nodes
\begin{scope} [every node/.style=graycirclestyle]
\node (A) at (0,1) {$i$};
\node (B) at (-1,0) {};
%\node (C) at (-0.3,0) {};
\node (D) at (1,0) {};
%\node (E) at (0.3,0) {};
\end{scope}
%edges
\begin{scope}[every node/.style={fill=white}, every edge/.style={draw,very thick}]
\path [->] (A) edge node {$\beta_1$} (B);
%\path [->] (A) edge node {$\beta_2$} (C);
\path [<-] (A) edge node {$\beta_3$} (D);
%\path [<-] (A) edge node {$\beta_4$} (E);
\end{scope}
\end{tikzpicture}}}
\qquad\underrightarrow{q_i\to\zeta_k q_i}\qquad
\vcenter{\hbox{\begin{tikzpicture}[scale=2]
\begin{scope} [every node/.style=graycirclestyle]
\node (A) at (0,1) {$i$};
\node (B) at (-1,0) {};
%\node (C) at (-0.3,0) {};
\node (D) at (1,0) {};
%\node (E) at (0.3,0) {};
\end{scope}
%edges
\begin{scope}[every node/.style={fill=white}, every edge/.style={draw,very thick}]
\path [->] (A) edge node {$\zeta_k\beta_1$} (B);
%\path [->] (A) edge node {$\zeta_k\beta_2$} (C);
\path [<-] (A) edge node {$\zeta_k^{-1}\beta_3$} (D);
%\path [<-] (A) edge node {$\zeta_k^{-1}\beta_4$} (E);
\end{scope}
\end{tikzpicture}}}
\end{equation}

\item Shifting a charge $q_i$ by (a multiple of) another charge $q_j$:
\begin{equation}
q_i \to q_i + \alpha q_j 
\end{equation}
where $\alpha$ is such that $|q_i+\alpha q_j|$ is equal to 1 or 2. On the diagram this corresponds to the addition, for every arrow connecting $q_j$ to $q_k\neq q_i$, of an arrow with the same label and orientation between $q_i$ and $q_k$. Schematically this amounts to the following move:
\begin{equation}	\label{eq:move2}
\vcenter{\hbox{\begin{tikzpicture}[scale=1.7]
\begin{scope} [every node/.style=graycirclestyle]
\node (A) at (-1,0) {$i$};
\node (B) at (1,0) {$j$};
\node (C) at (1,-2) {$k$};
\end{scope}
%edges
\begin{scope}[every node/.style={fill=white}, every edge/.style={draw,very thick}]
\path [->] (A) edge node {$\beta$} (B);
\path[->] (B) edge node {$\gamma$} (C);
\end{scope}
\end{tikzpicture}}}
\quad\underrightarrow{q_i\to q_i + \alpha q_j}\quad
\vcenter{\hbox{\begin{tikzpicture}[scale=1.7]
\begin{scope} [every node/.style=graycirclestyle]
\node (A) at (-1,0) {$i$};
\node (B) at (1,0) {$j$};
\node (C) at (1,-2) {$k$};
\end{scope}
%edges
\begin{scope}[every node/.style={fill=white}, every edge/.style={draw,very thick}]
\path [->] (A) edge node {$\beta + \alpha |q_j|$} (B);
\path[->] (B) edge node {$\gamma$} (C);
\path[->] (A) edge node {$\alpha \gamma$} (C);
\end{scope}
\end{tikzpicture}}}
\end{equation}
%ora la casistica diventa complicata, forse è meglio prima ridursi a H_ii = 2 e spiegare tutte le mosse lì, che è meno intricato
\end{itemize}

\begin{figure}
\includegraphics[width=\textwidth]{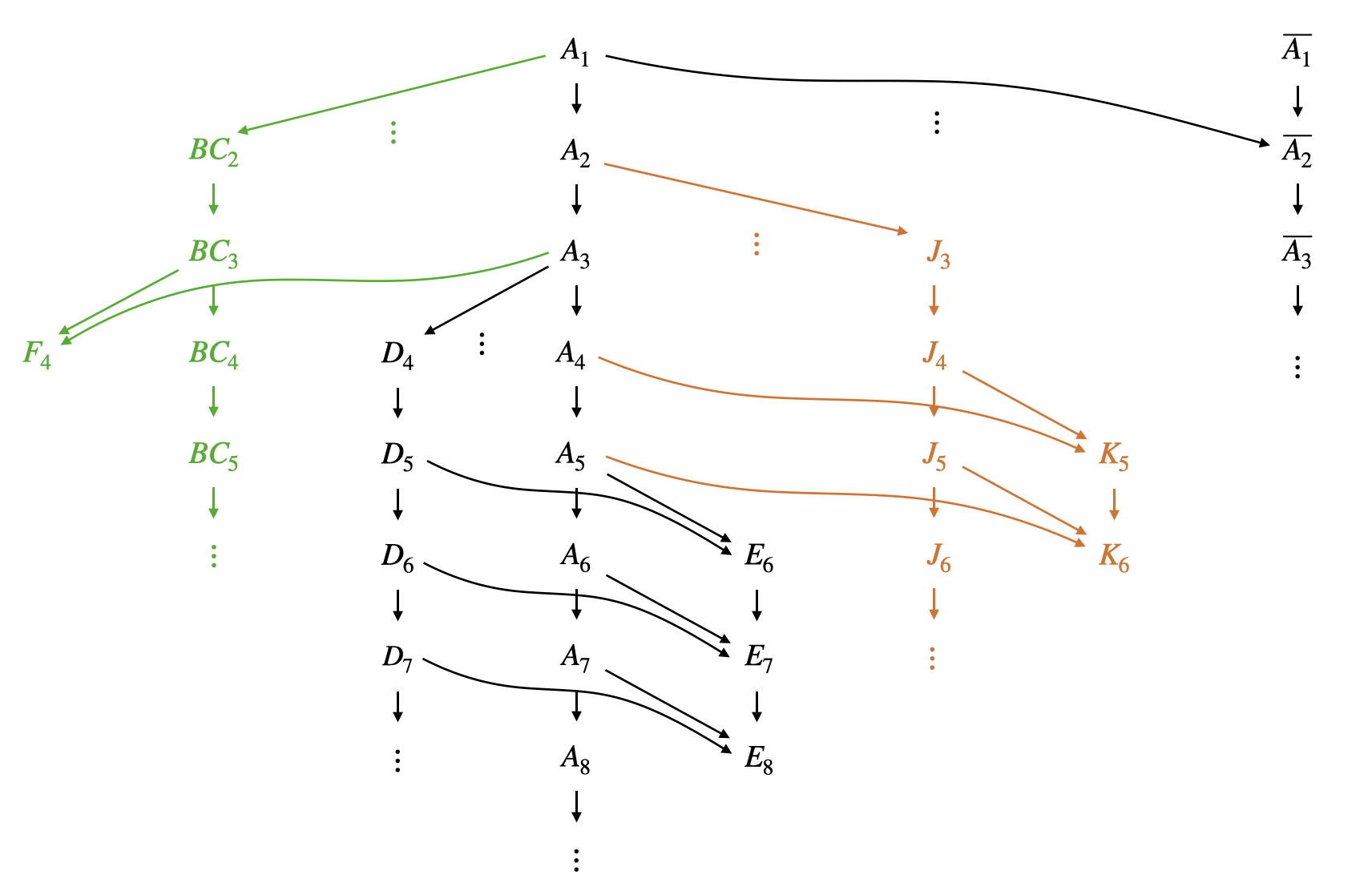}
\caption{Extension of charge lattices for $\NN=2$ SCFTs with $\varkappa \neq\{1,2\}$. The notation is the one used
 in Table \ref{tab:results}. 
 The arrows correspond to the possible ways a charge lattice of rank $r$ can be extended to rank $r+1$.
 The orange arrows and charge lattices only exist for $\varkappa \in \left\{3,\frac{2}{3}, 6, \frac{5}{6} \right\}$, the green arrows and charge lattices only exist for $\varkappa \in \left\{4, \frac{3}{4}\right\}$. }
\label{tab:inclusion_chart}
\end{figure}

We will show that by iteratively applying these basic changes of basis we are able to determine all inequivalent Dirac pairings. 
Our strategy for the classification is analogous to the one adopted in \cite{Cohen1976} and 
 proceeds as follows.
 We start at rank-1, where there are two inequivalent Dirac pairings represented by the 1-dimensional matrices $H=(1)$ and $H=(2)$. Than every rank-2 lattice can be obtained by starting with one of the rank-1 lattices and adding another node in the diagram with a general set of arrows between the new node and the old node. 
Borrowing the nomenclature of \cite{Cohen1976} we call these diagrams \textit{extensions} of the lower rank diagram.
From this list of diagrams we select the ones that correspond to positive-definite matrices and then we apply the basic moves described above to determine which diagrams are inequivalent.
Once the set of inequivalent diagrams is established we choose a representative for every equivalence class, add another node and repeat the procedure above to find all rank-3 charge lattices, and so on and so forth. 
The results are summarized in Figure \ref{tab:inclusion_chart}, which shows the various ways in which each charge lattice can be extended by adding a new node.

We will consider charge lattices of irreducible theories, that is rank-$r$ SCFTs that are not stacks of lower rank theories. Then the CB is not a Cartesian product of lower rank CBs and  there is at least one basis where the charge lattice is associated to a connected graph.
We will therefore classify all connected graphs. The most general charge lattice of a rank-$r$ SCFT with $\varkappa \neq \{1,2\}$ will be given by a disjoint union of connected diagrams.
In Sections \ref{subsec:rank23} through \ref{subsec:Jn} we classify the lattices that do not posses charges of length 1 (short charges).
Then in Section \ref{subsec:short} we will include short charges.

%The diagrams associated to these lattices have only double circles (long charges) and must maintain this property after changes of basis given by the moves above. One can show that in the case of lattices with only long charges the moves \eqref{eq:move1},   \eqref{eq:move2} map connected diagrams to connected diagrams. For irreducible theories, that is rank-$r$ theories that are not stacks of lower and theories, the CB is not a Cartesian product therefore in some basis of the charge lattice the corresponding diagram must be connected. 
%For lattices with only long charges this is the case in any basis and we can restrict ourselves to classifying connected 1diagrams. In the case of lattices with short charges the situation is more subtle\footnote{Namely when either $q_i$ or $q_j$ is short the move \eqref{eq:move2} can map connected diagrams to disconnected diagrams.} and we will tackle it in Section \ref{}.

% Pf(J) = det(H) 
As discussed in the introduction the order of the 1-form symmetry group of an $\NN=2$ theory is given by the Pfaffian of the Dirac pairing $J$. In the context of interest to us we can write the matrix representing $J$ using \eqref{eq:Hdef_formula}:
\begin{equation} 	\label{eq:Jtensor}
\begin{split}
J =& \frac{1}{\zeta- \overline{\zeta}} \left(
\left( \begin{array}{cc} 1 & \zeta \\ \overline{\zeta} & 1 \end{array} \right) \otimes H - 
\overline{\left( \begin{array}{cc} 1 & \zeta \\ \overline{\zeta} & 1 \end{array} \right) \otimes H} 
\right)
\\
=&
\frac{1}{\zeta- \overline{\zeta}} \left(
\left( \begin{array}{cc} 1 & \zeta \\ \overline{\zeta} & 1 \end{array} \right) \otimes H - 
\left(\left( \begin{array}{cc} 1 & \zeta \\ \overline{\zeta} & 1 \end{array} \right) \otimes H \right)^{T}
\right)
\end{split}
\end{equation}
Then the absolute value of the Pfaffian of $J$ is given by $\det(H)$. To show this notice that both $\det(H)$ and $|\Pf(J)|$ are holomorphic polynomials of the entries $H_{ij}$ of $H$. Furthermore when $H$ is real we can compute:
\begin{equation}
J = \left( \begin{array}{cc} 0 & H \\ -H &0 \end{array} \right)
\quad
\Rightarrow
\quad
\Pf(J) = (-1)^r \det(H)
\end{equation}
 $|\Pf(J)|$ and $\det(H)$ are both holomorphic polynomials in $H_{ij}$ and agree when $H_{ij} \in \mathbb{R}$, therefore they are equal for generic $H_{ij}$ as well.

%%%%%%%%%%%%%%%%%%%%%%%%%%%%%%%%%%%%%%%%%%%%%%%%%%%%%%%%%%%%%%%%%%%%%%%%%%%%%%%%
\subsection{Rank-2 and rank-3}
\label{subsec:rank23}

% table of H_ij
\begin{table} 
\begin{equation*}
\renewcommand{\arraystretch}{1.3}
\begin{array}{|c|c|c|}
\hline
 & k=4 & k=6 
 \\ \hline
 H_{ii}=H_{jj} = 2 & H_{ij}= i^n, i^n(1+i), \quad n=1,\dots,4 & H_{ij}=\omega^n, \omega^n(1+\omega), \quad n=1,\dots,6
 \\ \hline
  H_{ii}=1,H_{jj} = 2 & H_{ij}= i^n,  \quad n=1,\dots,4 & H_{ij}=\omega^n,  \quad n=1,\dots,6
   \\ \hline
  H_{ii}=H_{jj} = 1 & H_{ij}=0 & H_{ij}=0
  \\
  \hline
\end{array}
\renewcommand{\arraystretch}{1}
\end{equation*}
\caption{Allowed values of the off-diagonal elements of the Hermitian matrices $H$. Here and in the rest of the paper  $\omega=e^{\frac{\pi i }{3}}$.}
\label{tab:possibleHij}
\end{table}

\begin{comment}
% table of H_ij
\begin{table}
\begin{equation*}
\renewcommand{\arraystretch}{1.3}
\begin{array}{|c|c|c|}
\hline
 & k=4 & k=6 
 \\ \hline
 H_{ii}=H_{jj} = 2 & H_{ij}= \pm 1, \, \pm i, \; \pm1\pm i & H_{ij}=
 \begin{array}{l}
 \pm1,\; \pm\omega,\; \pm \omega^2,
 \\
 \pm(1+\omega),\; \pm\omega(1+\omega),\; \pm \omega^2(1+\omega)
 \end{array}
 \\ \hline
  H_{ii}=1,H_{jj} = 2 & H_{ij}=\pm 1, \, \pm i, & H_{ij}=\pm1,\; \pm\omega,\; \pm \omega^2
   \\ \hline
  H_{ii}=H_{jj} = 1 & H_{ij}=0 & H_{ij}=0
  \\
  \hline
\end{array}
\renewcommand{\arraystretch}{1}
\end{equation*}
\caption{Allowed values of the off-diagonal elements of the Hermitian matrices $H$. Here and in the rest of the paper  $\omega=e^{\frac{\pi i }{3}}$.}
\label{tab:possibleHij}
\end{table}
\end{comment}

At rank-1 there are two possible lattices corresponding to the single circle \singlenode and the double circle \doublenode.
In this Section we will focus on the classification of lattices with only long charges, and we will tackle the problem of lattices with short charges in Section \ref{subsec:short}.
All rank-2 lattices can be obtained by adding a new long node \doublenode, with a generic arrow between the first and second node such that the resulting diagram is connected. 
%One can show that the moves \eqref{eq:move1}, \eqref{eq:move2} map connected graphs with only long charges to connected graphs, therefore in this Section we only consider connected extensions.
The possible values for the label of the arrow are given by the possible values of $H_{ij}$ in Table \ref{tab:possibleHij}.
Therefore the most general rank-2 diagram is:
\begin{equation}
\vcenter{\hbox{\begin{tikzpicture}[scale=1.5]
%nodes
\begin{scope} [every node/.style=doublecirclestyle]
\node (A) at (-1,0) {$1$};
\node (B) at (1,0) {$2$};
\end{scope}
%edges
\begin{scope}[every node/.style={fill=white}, every edge/.style={draw,very thick}]
\path [->] (A) edge node {$\alpha$} (B);
\end{scope}
\end{tikzpicture}}}
\end{equation}
with:
\begin{equation}
\alpha = \left\{
\begin{array}{lll}
i^n, (i+1)i^n, \; &\qquad n=1,\dots,4 \qquad &\qquad k=4
\\
\omega^n, (1+\omega)\omega^n \; &\qquad n=1,\dots,6 \qquad &\qquad k=3,6
\end{array}\right.
\end{equation}
where $\omega=e^{\frac{\pi i }{3}}$.

We can multiply $q_1$ by a suitable power of $\zeta_k$ such that $\alpha$ is either $1, 1+i$ or $1+\omega$. Furthermore if $\alpha=1+\omega$ we can perform the charge of basis:

\begin{equation}	\label{eq:1pomega_short}
\vcenter{\hbox{\begin{tikzpicture}[scale=1.7]
%nodes
\begin{scope} [every node/.style=doublecirclestyle]
\node (A) at (-1,0) {$1$};
\node (B) at (1,0) {$2$};
\end{scope}
%edges
\begin{scope}[every node/.style={fill=white}, every edge/.style={draw,very thick}]
\path [->] (A) edge node {$1+\omega$} (B);
\end{scope}
\end{tikzpicture}}}
\quad\underrightarrow{q_1\to q_1 - q_2}\quad
\vcenter{\hbox{\begin{tikzpicture}[scale=1.7]
%nodes
\node[style={gray,circle,thick,draw=black}] (A) at (-1,0) {$1$};
\begin{scope} [every node/.style=doublecirclestyle]
\node (B) at (1,0) {$2$};
\end{scope}
%edges
\begin{scope}[every node/.style={fill=white}, every edge/.style={draw,very thick}]
\path [->] (A) edge node {$\omega$} (B);
\end{scope}
\end{tikzpicture}}}
\end{equation}
Therefore the corresponding lattice contains short charges, and we do not consider it in this Section. Similarly, for the rest of the section we will not consider diagrams containing arrow labeled by $(1+\omega)\omega^n$, because the corresponding lattice would contain a short charge.

Then the two inequivalent rank-2 charge lattices are given by the diagrams:

\begin{equation}
A_2 :
\begin{tikzpicture}[baseline=-3pt,,scale=1.5]
%nodes
\begin{scope} [every node/.style=doublecirclestyle]
\node (A) at (-1,0) {};
\node (B) at (1,0) {};
\end{scope}
%edges
\begin{scope}[every node/.style={fill=white}, every edge/.style={draw,very thick}]
\path [-] (A) edge  (B);
\end{scope}
\end{tikzpicture},
\qquad 
BC_2 :
\begin{tikzpicture}[baseline=-3pt,scale=1.5]
%nodes
\begin{scope} [every node/.style=doublecirclestyle]
\node (A) at (-1,0) {};
\node (B) at (1,0) {};
\end{scope}
%edges
\begin{scope}[every node/.style={fill=white}, every edge/.style={draw,very thick}]
\path [->] (A) edge node[above] {$1+i$} (B);
\end{scope}
\end{tikzpicture}
\end{equation}

That can be obtained as extensions of the $A_1$ diagram:
\begin{equation}
A_1 \longrightarrow A_2, BC_2
\end{equation}

Let us then proceed to study the rank-3 case by taking one of the two rank-2 diagrams and adding one node. Starting from the $A_2$ diagram the possible rank-3 diagrams are of the form:
\begin{equation}
\begin{tikzpicture}[baseline=-3pt,,scale=1.2]
%nodes
\begin{scope} [every node/.style=doublecirclestyle]
\node (A) at (-1,0) {};
\node (B) at (1,0) {};
\node (C) at (0, 1.7) {};
\end{scope}
%edges
\begin{scope}[every node/.style={fill=white}, every edge/.style={draw,very thick}]
\path [-] (A) edge  (B);
\path [->] (C) edge node[left] {$\alpha_1$}(A);
\path [->] (C) edge  node[right] {$\alpha_2$}(B);
\end{scope}
\end{tikzpicture}
\end{equation}
where the $\alpha_i$ are not both vanishing.
Up to multiplication of the new node by $\zeta_k$ and conjugation of the matrix $H$ the possible connected diagrams corresponding to positive definite matrices are:

\begin{equation}
\begin{gathered}
\begin{tikzpicture}[baseline=-3pt,,scale=1.2]
%nodes
\begin{scope} [every node/.style=doublecirclestyle]
\node (A) at (-1,0) {};
\node (B) at (1,0) {};
\node (C) at (0, 1.7) {};
\end{scope}
%edges
\begin{scope}[every node/.style={fill=white}, every edge/.style={draw,very thick}]
\path [-] (A) edge  (B);
\path [-] (C) edge (A);
%\path [->] (C) edge  node[right] {$\alpha_2$}(B);
\end{scope}
\end{tikzpicture},
\quad
\begin{tikzpicture}[baseline=-3pt,,scale=1.2]
%nodes
\begin{scope} [every node/.style=doublecirclestyle]
\node (A) at (-1,0) {};
\node (B) at (1,0) {};
\node (C) at (0, 1.7) {};
\end{scope}
%edges
\begin{scope}[every node/.style={fill=white}, every edge/.style={draw,very thick}]
\path [-] (A) edge  (B);
\path [-] (C) edge (A);
\path [-] (C) edge  node[right] {-1}(B);
\end{scope}
\end{tikzpicture},
\quad
\begin{tikzpicture}[baseline=-3pt,,scale=1.2]
%nodes
\begin{scope} [every node/.style=doublecirclestyle]
\node (A) at (-1,0) {};
\node (B) at (1,0) {};
\node (C) at (0, 1.7) {};
\end{scope}
%edges
\begin{scope}[every node/.style={fill=white}, every edge/.style={draw,very thick}]
\path [-] (A) edge  (B);
\path [-] (C) edge (A);
\path [->] (C) edge  node[right] {$i$}(B);
\end{scope}
\end{tikzpicture},
\\
\begin{tikzpicture}[baseline=-3pt,,scale=1.2]
%nodes
\begin{scope} [every node/.style=doublecirclestyle]
\node (A) at (-1,0) {};
\node (B) at (1,0) {};
\node (C) at (0, 1.7) {};
\end{scope}
%edges
\begin{scope}[every node/.style={fill=white}, every edge/.style={draw,very thick}]
\path [-] (A) edge  (B);
\path [-] (C) edge (A);
\path [->] (C) edge  node[right] {$1-i$}(B);
\end{scope}
\end{tikzpicture},
\quad
\begin{tikzpicture}[baseline=-3pt,,scale=1.2]
%nodes
\begin{scope} [every node/.style=doublecirclestyle]
\node (A) at (-1,0) {};
\node (B) at (1,0) {};
\node (C) at (0, 1.7) {};
\end{scope}
%edges
\begin{scope}[every node/.style={fill=white}, every edge/.style={draw,very thick}]
\path [-] (A) edge  (B);
%\path [-] (C) edge (A);
\path [->] (C) edge  node[right] {$1-i$}(B);
\end{scope}
\end{tikzpicture},
\quad
\begin{tikzpicture}[baseline=-3pt,,scale=1.2]
%nodes
\begin{scope} [every node/.style=doublecirclestyle]
\node (A) at (-1,0) {};
\node (B) at (1,0) {};
\node (C) at (0, 1.7) {};
\end{scope}
%edges
\begin{scope}[every node/.style={fill=white}, every edge/.style={draw,very thick}]
\path [-] (A) edge  (B);
\path [->] (C) edge node[left] {$-1+i$} (A);
\path [->] (C) edge  node[right] {$1-i$}(B);
\end{scope}
\end{tikzpicture},
\\
\begin{tikzpicture}[baseline=-3pt,,scale=1.2]
%nodes
\begin{scope} [every node/.style=doublecirclestyle]
\node (A) at (-1,0) {};
\node (B) at (1,0) {};
\node (C) at (0, 1.7) {};
\end{scope}
%edges
\begin{scope}[every node/.style={fill=white}, every edge/.style={draw,very thick}]
\path [-] (A) edge  (B);
\path [-] (C) edge (A);
\path [->] (C) edge  node[right] {$\omega$}(B);
\end{scope}
\end{tikzpicture}
\end{gathered}
\label{eq:list_rank3_diagrams}
\end{equation}

The first and second diagrams in \eqref{eq:list_rank3_diagrams} are equivalent as can be seen through the change of basis:
\begin{equation}
\vcenter{\hbox{\begin{tikzpicture}[baseline=-3pt,,scale=1.2]
%nodes
\begin{scope} [every node/.style=doublecirclestyle]
\node (A) at (-1,0) {$1$};
\node (B) at (1,0) {$2$};
\node (C) at (0, 1.7) {$3$};
\end{scope}
%edges
\begin{scope}[every node/.style={fill=white}, every edge/.style={draw,very thick}]
\path [-] (A) edge  (B);
\path [-] (C) edge (A);
\path [-] (C) edge  node[right] {-1}(B);
\end{scope}
\end{tikzpicture}}}
\quad\underrightarrow{q_1\to q_1+ q_2}\quad
\vcenter{\hbox{\begin{tikzpicture}[baseline=-3pt,,scale=1.2]
%nodes
\begin{scope} [every node/.style=doublecirclestyle]
\node (A) at (-1,0) {$1$};
\node (B) at (1,0) {$2$};
\node (C) at (0, 1.7) {$3$};
\end{scope}
%edges
\begin{scope}[every node/.style={fill=white}, every edge/.style={draw,very thick}]
\path [-] (A) edge  (B);
\path [-] (C) edge (A);
%\path [-] (C) edge  node[right] {-1}(B);
\end{scope}
\end{tikzpicture}}}
\end{equation}

Similarly the third, fourth, fifth and sixth diagrams of  \eqref{eq:list_rank3_diagrams} are all equivalent:

\begin{equation}
\vcenter{\hbox{\begin{tikzpicture}[baseline=-3pt,,scale=1.2]
%nodes
\begin{scope} [every node/.style=doublecirclestyle]
\node (A) at (-1,0) {$1$};
\node (B) at (1,0) {$2$};
\node (C) at (0, 1.7) {$3$};
\end{scope}
%edges
\begin{scope}[every node/.style={fill=white}, every edge/.style={draw,very thick}]
\path [-] (A) edge  (B);
\path [->] (C) edge node[left] {$-1+i$}(A);
\path [->] (C) edge  node[right] {$1-i$}(B);
\end{scope}
\end{tikzpicture}}}
\quad\underrightarrow{q_1\to q_1+ q_2}\quad
\vcenter{\hbox{\begin{tikzpicture}[baseline=-3pt,,scale=1.2]
%nodes
\begin{scope} [every node/.style=doublecirclestyle]
\node (A) at (-1,0) {$1$};
\node (B) at (1,0) {$2$};
\node (C) at (0, 1.7) {$3$};
\end{scope}
%edges
\begin{scope}[every node/.style={fill=white}, every edge/.style={draw,very thick}]
\path [-] (A) edge  (B);
%\path [-] (C) edge (A);
\path [->] (C) edge  node[right] {$1-i$}(B);
%\path [-] (C) edge  node[right] {-1}(B);
\end{scope}
\end{tikzpicture}}}
\end{equation}

\begin{equation}
\begin{gathered}
\vcenter{\hbox{\begin{tikzpicture}[baseline=-3pt,,scale=1.2]
%nodes
\begin{scope} [every node/.style=doublecirclestyle]
\node (1) at (-1,0) {$1$};
\node (2) at (1,0) {$2$};
\node (3) at (0, 1.7) {$3$};
\end{scope}
%edges
\begin{scope}[every node/.style={fill=white}, every edge/.style={draw,very thick}]
\path [-] (1) edge  (2);
\path [-] (3) edge (1);
\path [->] (3) edge  node[right] {$i$}(2);
\end{scope}
\end{tikzpicture}}}
\quad\underrightarrow{q_2\to q_2+ q_1}\quad
\vcenter{\hbox{\begin{tikzpicture}[baseline=-3pt,,scale=1.2]
%nodes
\begin{scope} [every node/.style=doublecirclestyle]
\node (1) at (-1,0) {$1$};
\node (2) at (1,0) {$2$};
\node (3) at (0, 1.7) {$3$};
\end{scope}
%edges
\begin{scope}[every node/.style={fill=white}, every edge/.style={draw,very thick}]
\path [-] (1) edge  (2);
\path [-] (3) edge (1);
\path [->] (3) edge  node[right] {$-1+i$}(2);
\end{scope}
\end{tikzpicture}}}
\longrightarrow
\\ 
\quad\underrightarrow{q_3\to q_3+ q_2}\quad
\vcenter{\hbox{\begin{tikzpicture}[baseline=-3pt,,scale=1.2]
%nodes
\begin{scope} [every node/.style=doublecirclestyle]
\node (1) at (-1,0) {$1$};
\node (2) at (1,0) {$2$};
\node (3) at (0, 1.7) {$3$};
\end{scope}
%edges
\begin{scope}[every node/.style={fill=white}, every edge/.style={draw,very thick}]
%\path [-] (1) edge  (2);
\path [-] (3) edge (1);
\path [->] (3) edge  node[right] {$1+i$}(2);
\end{scope}
\end{tikzpicture}}}
\end{gathered}
\end{equation}

Therefore we found that the $A_2$ diagram can be extended in three possible ways producing the $A_3$, $BC_3$ and $J_3$ diagrams. 
Performing analogous computations one can show that the $BC_2$ diagram can only be extended to the $BC_3$ diagram.
Then the inequivalent rank-3 charge lattices are:

\begin{equation}
\begin{gathered}
A_3 :
\begin{tikzpicture}[baseline=-3pt,,scale=1]
%nodes
\begin{scope} [every node/.style=doublecirclestyle]
\node (A) at (-1,0) {};
\node (B) at (1,0) {};
\node (C) at (3,0) {};
\end{scope}
%edges
\begin{scope}[every node/.style={fill=white,circle}, every edge/.style={draw,very thick}]
\path [-] (A) edge  (B);
\path [-] (B) edge  (C);
\end{scope}
\end{tikzpicture},
\qquad 
BC_3 :
\begin{tikzpicture}[baseline=-3pt,scale=1]
%nodes
\begin{scope} [every node/.style=doublecirclestyle]
\node (A) at (-1,0) {};
\node (B) at (1,0) {};
\node (C) at (3,0) {};
\end{scope}
%edges
\begin{scope}[every node/.style={fill=white,circle}, every edge/.style={draw,very thick}]
\path [->] (A) edge node[above] {$1+i$} (B);
\path [-] (B) edge  (C);
\end{scope}
\end{tikzpicture},
\vspace{0.5cm}\\ 
J_3:
\vcenter{\hbox{
\begin{tikzpicture}[baseline=-3pt,,scale=1]
%nodes
\begin{scope} [every node/.style=doublecirclestyle]
\node (A) at (-1,0) {};
\node (B) at (1,0) {};
\node (C) at (0, 1.7) {};
\end{scope}
%edges
\begin{scope}[every node/.style={fill=white,circle}, every edge/.style={draw,very thick}]
\path [-] (A) edge  (B);
\path [-] (C) edge (A);
\path [->] (C) edge  node[right] {$\omega$}(B);
\end{scope}
\end{tikzpicture}}}
\end{gathered}
\end{equation}

That are obtained as extensions of the $A_2$ and $BC_2$ diagrams:
\begin{equation}
A_2 \longrightarrow A_3, BC_3, J_3,
\quad
BC_2\longrightarrow BC_3
\end{equation}

%%%%%%%%%%%%%%%%%%%%%%%%%%%%%%%%%%%%%%%%%%%%%%%%%%%%%%%%%%%%%%%%%%%%%%%%%%%%%%%%
\section{Classification of charge lattices for $\varkappa \neq \{1,2\}$}
\label{sec:3}

The procedure outlined in Section \ref{sec:2} can be carried out for higher ranks, thought computing all possible extensions and identifying the equivalent diagrams becomes increasingly demanding. 
We expect that diagrams shaped as ordinary Dynkin diagrams will appear in the classification. A property shared by all Dynkin diagrams at rank-$r$ is the presence is a $A_n$ subgraph with $n$ of order $\mathcal{O}(r)$. Then in any extension of these diagrams most of the arrows will be directed from the new node $q_0$ to one of the nodes of the $A_n$ subgraph.
It is then useful to simplify as much as possible the subgraph containing the new node and the $A_n$ subgraph.

%%%%%%%%%%%%%%%%%
\subsection{Extensions of $A_r$}
\label{subsec:ext_Ar}
\def\scaledoomtitans{0.7}

In this Section we consider the possible extensions of the $A_r$ diagram. 
We will simplify the extensions by only considering charges of basis of the form $q_0 \to q_o + \alpha q_i$ for some $i$. These changes of basis modify the arrows between the node $q_0$ and the other nodes but they do not modify, add or remove any arrow between the other nodes $q_i$ for $i\geq 1$. 
This implies that the same changes of basis can be applied when
considering the extensions of a diagram $D$ with an $A_r$ subgraph. The changes of basis $q_0 \to q_o + \alpha q_i$ will not affect the edges od $D$, significantly reducing the amount of extensions of $D$ that must be considered. 

The most generic extension of $A_r$ is a diagram with the following form:
\begin{equation}	\label{eq:An_extension_gen}
\sdiagramAloop[scale=1.5](0)(\alpha_{j-1},\alpha_{j},\alpha_{j+1},\alpha_{k-1},\alpha_{k},\alpha_{k+1})
\end{equation}

where at least one of the $\alpha_i$ is non-vanishing.
Let us first consider the case where all $\alpha_i$ are either $1$, $-1$ or $0$, and denote as $\{j_i\}$ the set of nodes that are connected to the new node $q_0$, ordered from left to right. Then the subgraph containing $q_0$, $q_{j_i}, q_{j_i+1}, \dots q_{j_{i+1}}$ is a cycle and positive definiteness requires $\alpha_{j_i} = 1, \alpha_{j_i} = -1$ or viceversa.
If the node to the left of $j_i$ is not connected to $q_0$ we can “move" this triangle to the left via the following change of basis.
\begin{equation}	\label{eq:trianglemoveleft}
   \makebox[\displaywidth]{$\displaystyle
\sdiagramAloop[scale=\scaledoomtitans](0)(0,1,0,0,-1,0)
\quad\underrightarrow{q_0\to q_0 - \sum_{\ell=j}^{k+1} q_\ell}\quad
\sdiagramAloop[scale=\scaledoomtitans](0)(1,0,0,-1,0,0)
$}
\end{equation}
If the node to the left of $j_{i}$ is connected to $q_0$ or if $j_{i}$ is the leftmost node of the $A_n$ diagram, the same change of basis produces:
\begin{equation}	\label{eq:trianglemovecollision}
   \makebox[\displaywidth]{$\displaystyle
\sdiagramAloop[scale=\scaledoomtitans](0)(-1,1,0,0,-1,0)
\quad\underrightarrow{q_0\to q_0 - \sum_{\ell=j_{i}}^{j_{i+1}-1} q_\ell}\quad
\sdiagramAloop[scale=\scaledoomtitans](0)(0,0,0,-1,0,0)
$}
\end{equation}

\begin{equation}	\label{eq:trianglemovefinishAn}
   \makebox[\displaywidth]{$\displaystyle
\sdiagramAleft[scale=\scaledoomtitans](0)(1,0,0,0,-1,0)
\quad\underrightarrow{q_0\to q_0 - \sum_{\ell=1}^{j_{i+1}-1} q_\ell}\quad
\sdiagramAleft[scale=\scaledoomtitans](0)(0,0,0,-1,0,0)
$}
\end{equation}

In both cases the resulting diagram has less arrows connecting the new node with respect to the diagram we started with. We can iteratively apply the moves \eqref{eq:trianglemoveleft}, \eqref{eq:trianglemovecollision} and \eqref{eq:trianglemovefinishAn} to reduce the any diagram until there is only one 
node, say $k$, connected to the new node $q_0$.
Furthermore if the node $k$ connected to $q_0$ is on the right half of $A_n$, namely $k> \frac{n}{2}$, one can perform similar changes of basis to move it to $n-k+1$, which is on the left half of $A_n$.
Positivity of the determinant imposes $k=1,2$ or $3$ producing the diagrams $A_{n+1}$, $D_{n+1}$ and $E_{6,7,8}$ respectively.

%%%%%%%%%
\subsubsection{Case $\varkappa \in \left\{4,\frac{3}{4} \right\}$}
Let us now consider the case of generic values of $\alpha_i$. We start from the case $k=4$, where the possible value of $\alpha_i$ are:
\begin{equation}
\alpha_i = \pm1, \;\pm i, \;\pm 1 \pm i
\end{equation}
and write each $\alpha_i$ isolating the factors $\pm1$ and the factors $\pm i$:
\begin{equation}
\alpha_i = \beta_i + \gamma_i,
\qquad
\beta_i = \pm1, \; \gamma_i = \pm i
\end{equation}

Now we can apply the simplifying procedure based on the moves \eqref{eq:trianglemoveleft}, \eqref{eq:trianglemovecollision} and \eqref{eq:trianglemovefinishAn} as if there were only $\beta_i$ edges. This is due to the fact that the $\gamma_i$ arrows are not affected by the moves above:
\begin{equation}
   \makebox[\displaywidth]{$\displaystyle
\begin{gathered}
	\sdiagramAloop[scale=\scaledoomtitans](0)(0,0,1,0,i,-1)
	\quad\underrightarrow{q_0\to q_0 - \sum_{\ell=j_i+1}^{j_{i+1}} q_\ell}\quad
	\sdiagramAloop[scale=\scaledoomtitans](0)(0,1,0,0,i-1,0)
\\
	\quad\underrightarrow{q_0\to q_0 - \sum_{\ell=j_i}^{j_{i+1}-1} q_\ell}\quad
	\sdiagramAloop[scale=\scaledoomtitans](0)(1,0,0,-1,i,0)
\end{gathered}
$}
\end{equation}

After running the simplifying procedure described above the diagram will contain a single $\beta_i$ arrow and the original $\gamma_i$ arrows. Then we can multiply $q_0$ by $i$, so that all $\gamma_i$ became either $1$ or $-1$ and run the simplifying procedure on the $\gamma_i$. The resulting diagram has at most one $\beta$ arrow connected to, say, the $k_1$ node and one $\gamma$ arrow connected to the $k_2$ node. 
Both $k_1$ and $k_2$ can be moved to sit in the left half of $A_n$ and without loss of generality we can take $k_1 \leq k_2$:
\begin{equation}
\sdiagramAreduced[scale=1](0)(\beta, \gamma)(k_1,k_2),
\quad
k_1 \leq k_2 \leq \left\lceil \frac{n}{2} \right\rceil
\end{equation}

If either $\beta$ or $\gamma$ is $0$ the resulting diagram is equivalent to $A_{n+1}$, $D_{n+1}$ or $E_{6,7,8}$.
%for generic $n$ or $E_{n+1}$ for $n+1=6,7,8$, because other diagrams would have non-positive determinant. 
%Furthermore, through manipulations similar to the ones described above, the diagram can be put in a form such that the only node connected to $q_0$ is $1,2$ or $3$ for $A_n, D_n$ or $E_n$ respectively.
If $k_1=k_2=k$ the only diagram with positive determinant is given by $k=1$ or $k=n$, giving the $BC_{n+1}$ diagram either way.

When both $\beta$ and $\gamma$ are non-vanishing and $k_1 \neq k_2$ we can multiply $q_0$ by $-\overline{\beta}$ and obtain the diagram:
\begin{equation}	\label{eq:singleloopextension}
\extensiondiagram[A_n]_{k_1,k_2}^{\delta} = \sdiagramAreduced[scale=1](0)(-1, \delta)(k_1,k_2),
\quad
k_1 < k_2 \leq \left\lceil \frac{n}{2} \right\rceil
\end{equation}

where $\delta=-\overline{\beta} \gamma$.
The allowed diagrams are those with positive determinant.
The determinant of $\extensiondiagram[A]_{k_1,k_2}^{\delta} $ can be computed by induction and for the case of interest here, namely with $|\delta|=1$, the result is:
\begin{equation} 	\label{eq:singleloopextension_det}
\det(\extensiondiagram[A_n]_{k_1,k_2}^{\delta} ) = k_1^2 + k_2^2 - (n+1)(k_1+k_2-2) + 2 k_1 (n-1-k_2) \Re[\delta]
\end{equation}

For $\delta=\pm i$ this reduces to:
\begin{equation}
\det(\extensiondiagram[A_n]_{k_1,k_2}^{\pm i} ) = k_1^2 + k_2^2 - (n+1)(k_1+k_2-2) 
\end{equation}

In the range $k_1 < k_2 \leq \left\lceil \frac{n}{2} \right\rceil$ the only diagram $\extensiondiagram[A_n]_{k_1,k_2}^{\pm i}$ with positive determinant is obtained with $n=3$, $k_1 = 1$ and $k_2 = 2$.
\begin{equation}
\extensiondiagram[A_3]_{1,2}^{\pm i} =
\vcenter{\hbox{\begin{tikzpicture}[scale=1]
%nodes
\begin{scope} [every node/.style=doublecirclestyle]
\node (A) at (-1,0) {};
\node (B) at (1,0) {};
\node (C) at (3,0) {};
\node (D) at (0,1.7) {};
\end{scope}
%edges
\begin{scope}[every node/.style={fill=white}, every edge/.style={draw,very thick}]
\path [-] (A) edge  (B);
\path [-] (B) edge  (C);
\arrowedge[\pm i](D, B)()
\arrowedge[-1](D, A)()
\end{scope}
\end{tikzpicture}}},
\qquad
\det(\extensiondiagram[A_3]_{1,2}^{\pm i}) = 1
\end{equation}
%In Appendix \ref{app} we show that $\extensiondiagram[A_3]_{1,2}^{\pm i}$ is equivalent to the $F_4$ diagram. 
which is equivalent to $F_4$, as can be seen from the change of basis:
\begin{equation}
\vcenter{\hbox{\begin{tikzpicture}[scale=.95]
%nodes
\begin{scope} [every node/.style=doublecirclestyle]
\node (A) at (-1,0) {1};
\node (B) at (1,0) {2};
\node (C) at (3,0) {3};
\node (D) at (0,1.7) {0};
\end{scope}
%edges
\begin{scope}[every node/.style={fill=white}, every edge/.style={draw,very thick}]
\path [-] (A) edge  (B);
\path [-] (B) edge  (C);
\arrowedge[\pm i](D, B)()
\arrowedge[-1](D, A)()
\end{scope}
\end{tikzpicture}}}
	\quad\underrightarrow{q_2\to q_2 + q_1 \mp i q_0}\quad
\vcenter{\hbox{\begin{tikzpicture}[scale=.95]
%nodes
\begin{scope} [every node/.style=doublecirclestyle]
\node (A) at (-1,0) {1};
\node (B) at (1,0) {2};
\node (C) at (3,0) {3};
\node (D) at (-3,0) {0};
\end{scope}
%edges
\begin{scope}[every node/.style={fill=white}, every edge/.style={draw,very thick}]
\path [-] (D) edge  (A);
\path [->] (A) edge node[above=0.2] { $1+i$} (B);
\arrowedge[-1](B, C)()
\end{scope}
\end{tikzpicture}}}
\end{equation}

To summarize we found that the possible charge lattices obtained as extensions of the $A_n$ diagram with $\varkappa \in \left\{4,\frac{3}{4} \right\}$ are the following:
 \begin{equation}
 A_n \; \longrightarrow \;
A_{n+1}, \; D_{n+1}, \; E_{6,7,8},\; B_{n+1}, \; F_4
\qquad\qquad \varkappa \in \left\{4,\frac{3}{4} \right\}
 \end{equation}

%%%%%%%%%%%%%%%%%%
\subsubsection{Case $\varkappa \in \left\{3,\frac{2}{3}, 6, \frac{5}{6} \right\}$}
\def\scaledoubleloopdiag{1.7}
Let us now consider the case $k=3,6$, where the most generic extension of $A_n$ is given by the diagram \eqref{eq:An_extension_gen} with $\alpha_i = \pm 1, \pm\omega, \pm \omega^2$.
Let us introduce some notation that will be useful in reducing these extensions.
Given two consecutive nodes $j_i$ and $j_{i+1}$ connected to the new node $q_0$ we denote the triangle-shaped subgraph containing the nodes $j_i, j_i + 1, \dots, j_{i+1}$ and $0$ as a loop $L_{j_i,j_{i+1}}$ of weight $W[L_{j_i,j_{i+1}}]=\overline{\alpha_i} \alpha_{i+1}$.
\begin{equation}
L_{j_i,j_{i+1}} = 
\sdiagramLoop[scale=1](0)(\alpha_i,\alpha_{i+1})(j_i,j_{i+1}),
\qquad
W[L_{j_i,j_{i+1}}]=\overline{\alpha_i} \alpha_{i+1}
\end{equation}

Similarly we denote a subgraph delimited by three consecutive nodes connected to $q_0$ as a double loop $DL_{j_i,j_{i+1},j_{i+2}}$:
% and call the number of nodes in this subgraph its rank:
\begin{equation}
DL_{j_i,j_{i+1},j_{i+1}} = 
\sdiagramDoubleLoop[scale=1](0)(\alpha_i,\alpha_{i+1}, \alpha_{i+2})(j_i,j_{i+2},{j_{i+2}}),
%\qquad
%\text{rank}(DL_{j_i,j_{i+1},j_{i+2}}) = j_{i+1}-j_i +2
\end{equation}
which can be considered as two consecutive single loops with weights $\overline{\alpha_i} \alpha_{i+1}$ and $\overline{\alpha_{i+1}} \alpha_{i+2}$.

First we notice that if there is any loop with weight $-\omega^2$ or $-\overline{\omega^2}$ then the charge lattice contains a short charge. Without loss of generality we set $\alpha_i=-1$ and perform the change of basis:
\begin{equation}
\sdiagramLoop[scale=1](0)(-1,{\omega^2,\overline{\omega^2}})(j_i,j_{i+1})
	\quad\underrightarrow{q_0\to q_0 + \sum_{\ell=j_i}^{j_{i+1}-1} q_\ell}\quad
\sdiagramLoop[scale=1](0)(0,{\omega^2-1,\overline{\omega^2}}-1)(j_i,j_{i+1})
\end{equation}
and using \eqref{eq:1pomega_short} one finds a short charge. 
Furthermore one can check that a loop with weight $1$ has negative determinant.
For this reason in this section we only consider extensions containing loops with weight $-1,-\omega$ or $ -\overline{\omega}$.

As a first step we want to show that every extension is equivalent to an extension where at most two nodes are connected to the new node $q_0$, similarly to the $k=4$ case. 
We consider an extension with at least three nodes connected to $q_0$ and study one of the double loops. 
Without loss of generality we can set $\alpha_i=1$ and we relabel the nodes as follows:
\begin{equation}
DL_{1,k,r-1}^{\alpha,\beta} = \sdiagramDoubleLoop[scale=1](0)(-1,\alpha, \beta)(1,k,r-1)
\end{equation}
where $r$ is the rank of the double loop. The determinant of this double loop is:
\begin{equation}
\det(DL_{1,k,r-1}^{\alpha,\beta})
=
2 + 2\Re[\alpha] + 2 \Re[\beta] (r-k) -k(r-k) -2k \Re\left[ \frac{\alpha}{\beta} \right]
\end{equation}

We are interested in double loops with positive determinant that only include loops of weight $-1,-\omega$ and $-\overline{\omega}$. If both loops in the double loop have weight $\pm 1$ the double loop can be reduced to a single loop by the moves  \eqref{eq:trianglemoveleft}, \eqref{eq:trianglemovecollision} and \eqref{eq:trianglemovefinishAn} described above, therefore we focus on double loops with at least one loop with weight $-\omega$ or $-\overline{\omega}$.
Up to complex conjugation and reflection the relevant double loops with positive determinant are:
\begin{equation}	\label{eq:doubleloops1}
DL_{1,2,r-1}^{1,-\omega} = \sdiagramDoubleLoop[scale=1](0)(-1,1, -\omega)(1,2,r-1),
\qquad
 \forall r
\end{equation}

\begin{equation}	\label{eq:doubleloops2}
DL_{1,r-2,r-1}^{1,-\omega} = \sdiagramDoubleLoop[scale=1](0)(-1,1, -\omega)(1,r-2,r-1), 
\qquad 
r=5,6
\end{equation}

\begin{equation}	\label{eq:doubleloops3}
DL_{1,r-2,r-1}^{\omega,-\omega^2} = \sdiagramDoubleLoop[scale=1](0)(-1,\omega, -\omega^2)(1,r-2,r-1), 
\qquad 
r=4,5,6
\end{equation}

For the first two cases \eqref{eq:doubleloops2}, \eqref{eq:doubleloops2} the double loop can be reduced to a single loop as follows:
\begin{equation}
\sdiagramDoubleLoop[scale=1](0)(-1,1, -\omega)(1,2,r-1)
	\quad\underrightarrow{q_0\to q_0 - \sum_{\ell=2}^{r-2} q_\ell}\quad
\sdiagramDoubleLoop[scale=1](0)(0,-1, 1-\omega)(1,r-2,r-1)
\end{equation}

\begin{equation}
\sdiagramDoubleLoop[scale=1](0)(-1,1, -\omega)(1,r-2,r-1)
	\quad\underrightarrow{q_0\to q_0 - \sum_{\ell=2}^{r-2} q_\ell}\quad
\sdiagramDoubleLoop[scale=1](0)(0,-1,1 -\omega)(1,2,r-1)
\end{equation}

Therefore any double loop, except for those in \eqref{eq:doubleloops3}, can be reduced to a single loop by one of the moves described above. By applying those moves iteratively any extension of $A_n$ can be brought in a form which has either no loops, a single loop or contains double loops of the form \eqref{eq:doubleloops3}.
In the latter case, which can also be described as an extension where all loops have weight $-\omega$, consider the leftmost double loop. We can move the whole double loop to the left by the change of basis:
\begin{equation}
\begin{array}{l}
\vcenter{\hbox{\begin{tikzpicture}[scale=\scaledoubleloopdiag]
%nodes
%left and right ghost nodes
	\node (left) at (-2.5,0) {};
	\node (right) at (2.5,0) {};
	
%nodes
\begin{scope} [every node/.style=doublecirclestyle] 
	\node[label=below:$k-1$,double,circle,thick] (1) at (-2,0) {};
	\node[label=below:$k$,double,circle,thick]  (2) at (-1,0) {};
	\node[label=below:$j-1$,double,circle,thick]  (center) at (0,0) {};
	\node[label=below:$j$,double,circle,thick]  (3) at (1,0) {};
	\node[label=below:$j+1$,double,circle,thick]  (4) at (2,0) {};
	%new node
	\node[double,circle,thick] (new) at (0,1) {0};
\end{scope}

%edges
\begin{scope}[every node/.style={fill=white}, every edge/.style={draw,very thick}]
	\path [-] (1) edge (2);
	\path [dashed] (2) edge (center);
	\path [-] (center) edge (3);
	\path [-] (3) edge (4);
%left and right dashed lines
	\path[dashed] (left) edge (1);
	\path[dashed] (4) edge (right);
\end{scope}

%new arrows
\begin{scope}[every node/.style={fill=white}, every edge/.style={draw,very thick}]
	\arrowedge[-1](new,2)()
	\arrowedge[\omega](new,3)()
	\arrowedge[-\omega^2](new,4)()
\end{scope}
\end{tikzpicture}}}
\\ \underrightarrow{q_0\to q_0 + \sum_{\ell=k}^{j-1} q_\ell + \omega q_j}\quad	
\vcenter{\hbox{\begin{tikzpicture}[scale=\scaledoubleloopdiag]
%nodes
%left and right ghost nodes
	\node (left) at (-2.5,0) {};
	\node (right) at (2.5,0) {};
	
%nodes
\begin{scope} [every node/.style=doublecirclestyle] 
	\node[label=below:$k-1$,double,circle,thick] (1) at (-2,0) {};
	\node[label=below:$k$,double,circle,thick]  (2) at (-1,0) {};
	\node[label=below:$j-1$,double,circle,thick]  (center) at (0,0) {};
	\node[label=below:$j$,double,circle,thick]  (3) at (1,0) {};
	\node[label=below:$j+1$,double,circle,thick]  (4) at (2,0) {};
	%new node
	\node[double,circle,thick] (new) at (0,1) {0};
\end{scope}

%edges
\begin{scope}[every node/.style={fill=white}, every edge/.style={draw,very thick}]
	\path [-] (1) edge (2);
	\path [dashed] (2) edge (center);
	\path [-] (center) edge (3);
	\path [-] (3) edge (4);
%left and right dashed lines
	\path[dashed] (left) edge (1);
	\path[dashed] (4) edge (right);
\end{scope}

%new arrows
\begin{scope}[every node/.style={fill=white}, every edge/.style={draw,very thick}]
	\arrowedge[-1](new,1)()
	\arrowedge[\omega](new,center)()
	\arrowedge[-\omega^2](new,3)()
\end{scope}
\end{tikzpicture}}}
\end{array}
\end{equation}

We can move this double loop until we reach the leftmost node of $A_n$ and then simplify it with the change of basis:
\begin{equation}
\begin{array}{l}
\vcenter{\hbox{\begin{tikzpicture}[scale=\scaledoubleloopdiag]
%nodes
%left and right ghost nodes
	\node (right) at (2.5,0) {};
	
%nodes
\begin{scope} [every node/.style=doublecirclestyle] 
	\node[label=below:$1$,double,circle,thick]  (2) at (-1,0) {};
	\node[label=below:$j-1$,double,circle,thick]  (center) at (0,0) {};
	\node[label=below:$j$,double,circle,thick]  (3) at (1,0) {};
	\node[label=below:$j+1$,double,circle,thick]  (4) at (2,0) {};
	%new node
	\node[double,circle,thick] (new) at (0,1) {0};
\end{scope}

%edges
\begin{scope}[every node/.style={fill=white}, every edge/.style={draw,very thick}]
	\path [dashed] (2) edge (center);
	\path [-] (center) edge (3);
	\path [-] (3) edge (4);
%left and right dashed lines
	\path[dashed] (4) edge (right);
\end{scope}

%new arrows
\begin{scope}[every node/.style={fill=white}, every edge/.style={draw,very thick}]
	\arrowedge[-1](new,2)()
	\arrowedge[\omega](new,3)()
	\arrowedge[-\omega^2](new,4)()
\end{scope}
\end{tikzpicture}}}
\\ \underrightarrow{q_0\to q_0 + \sum_{\ell=1}^{j-1} q_\ell + \omega q_j}\quad	
\vcenter{\hbox{\begin{tikzpicture}[scale=\scaledoubleloopdiag]
%nodes
%left and right ghost nodes
	\node (right) at (2.5,0) {};
	
%nodes
\begin{scope} [every node/.style=doublecirclestyle] 
	\node[label=below:$1$,double,circle,thick]  (2) at (-1,0) {};
	\node[label=below:$j-1$,double,circle,thick]  (center) at (0,0) {};
	\node[label=below:$j$,double,circle,thick]  (3) at (1,0) {};
	\node[label=below:$j+1$,double,circle,thick]  (4) at (2,0) {};
	%new node
	\node[double,circle,thick] (new) at (0,1) {0};
\end{scope}

%edges
\begin{scope}[every node/.style={fill=white}, every edge/.style={draw,very thick}]
	\path [dashed] (2) edge (center);
	\path [-] (center) edge (3);
	\path [-] (3) edge (4);
%left and right dashed lines
	\path[dashed] (4) edge (right);
\end{scope}

%new arrows
\begin{scope}[every node/.style={fill=white}, every edge/.style={draw,very thick}]
	\arrowedge[\omega](new,center)()
	\arrowedge[-\omega^2](new,3)()
\end{scope}
\end{tikzpicture}}}
\end{array}
\end{equation}

This implies that any double loop can be reduced to a single loop with a suitable change of basis, therefore one can simplify any extension of $A_n$ to a diagram that contains at most two nodes connected to the new node $q_0$. If there is only one such connection the resulting diagram is an $ADE$ diagram while if there are two connection the extension is given by \eqref{eq:singleloopextension} with $\beta=\omega$:

\begin{equation}	\label{eq:oneloopext_k6}
\extensiondiagram[A_n]_{k_1,k_2}^{\omega} = \sdiagramAreduced[scale=1](0)(-1, \omega)(k_1,k_2)
\end{equation}

The determinant of this diagram is given by \eqref{eq:singleloopextension_det} with $\delta=\omega=e^{\frac{\pi i}{3}}$:
\begin{equation} 
\det(\extensiondiagram[A_n]_{k_1,k_2}^{\omega} ) = k_1^2 + k_2^2 - (n+1)(k_2-2) -k_1 k_2
\end{equation}
The diagrams with positive determinant are:
\begin{equation}
\extensiondiagram[A_n]_{1,2}^{\omega}, \;
\extensiondiagram[A_n]_{n-1,n}^{\omega}, \;
\extensiondiagram[A_n]_{1,n}^{\omega} 
\qquad \forall n
\end{equation}
and:
\begin{equation}
\begin{split}
&\extensiondiagram[A_4]_{2,3}^{\omega},
\\
&\extensiondiagram[A_5]_{2,3}^{\omega}, \;
\extensiondiagram[A_5]_{3,4}^{\omega}, \;
\end{split}
\end{equation}

The diagram $\extensiondiagram[A_n]_{n-1,n}^{\omega}$ is equivalent to the diagram $\extensiondiagram[A_n]_{1,n}^{\omega} $ as can be seen from the change of basis $q_0 \to q_0 + \sum_{\ell=1}^{n-1} q_\ell$. In turn the diagram $\extensiondiagram[A_n]_{1,n}^{\omega} $ is equivalent to $\extensiondiagram[A_n]_{1,2}^{\omega}$ through  $q_0 \to q_0 + \omega^2 \sum_{\ell=2}^{n-2} q_\ell$. 
Similarly one can show that $\extensiondiagram[A_5]_{2,3}^{\omega}$ and $\extensiondiagram[A_5]_{3,4}^{\omega}$ are equivalent.
Therefore the extensions of the form \eqref{eq:oneloopext_k6} are equivalent to one of the following:

\begin{equation}	\label{eq:axtensionA_Jleft}
\extensiondiagram[A_n]_{1,2}^{\omega} = J_{n+1} = \quad
\vcenter{\hbox{\begin{tikzpicture}[]
	
%nodes
\begin{scope} [every node/.style=doublecirclestyle] 
	\node[label=below:$1$,	double,circle,thick] (jp1) at (-1,0) {};
	\node[label=below:$2$,	double,circle,thick]  (km1) at (1,0) {};
	\node[label=below:$n$,	double,circle,thick]  (kp1) at (3,0) {};
	%new node
	\node[double,circle,thick] (new) at (0,1.7) {0};
\end{scope}

%edges
\begin{scope}[every node/.style={fill=white}, every edge/.style={draw,very thick}]
	\path [-] (jp1) edge (km1);
	\path [dashed] (km1) edge (kp1);
\end{scope}

%new arrows
\begin{scope}[every node/.style={fill=white}, every edge/.style={draw,very thick}]
	\arrowedge[-1](new,jp1)()
	\arrowedge[\omega](new,km1)()
\end{scope}
\end{tikzpicture}}}
\end{equation}

\begin{equation}
\extensiondiagram[A_4]_{2,3}^{\omega} =  K_5 = \quad \diagramKfive[](,2,3,,0)
\end{equation}

\begin{equation}
\extensiondiagram[A_5]_{2,3}^{\omega} =  K_6 =  \quad \diagramKsix[](,2,3,,,0)
\end{equation}

which, together with the extensions with only one node connected to $q_0$ discussed above, provide all possible extensions of the $A_n$ diagrams with $\varkappa \in \left\{3,\frac{2}{3}, 6, \frac{5}{6} \right\}$:
 \begin{equation}
 A_n \; \longrightarrow \;
A_{n+1}, D_{n+1}, E_{6,7,8},J_{n+1}, K_{5,6}
\qquad\qquad \varkappa \in \left\{3,\frac{2}{3}, 6, \frac{5}{6} \right\}
 \end{equation}

This concludes the analysis of the extensions of the $A_n$ diagrams. To summarize we found that the possible extensions of the $A_n$ diagrams are:
\begin{equation}	\label{eq:extensions_An}
A_n \quad \longrightarrow \quad
\left \{
\begin{array}{ll}
A_{n+1}, D_{n+1}, E_{6,7,8}, \qquad\qquad & \varkappa \neq\{1,2\}
\\
B_{n+1}, F_{4} &\varkappa \in \left\{4,\frac{3}{4} \right\}
\\
J_{n+1}, K_{5,6} &\varkappa \in \left\{3,\frac{2}{3}, 6, \frac{5}{6} \right\}
\end{array}
\right.
\end{equation}

As anticipated in Figure \ref{tab:inclusion_chart} these are all the charge lattices that only include long charges. 
In order to complete the classification of charge lattices for $\NN=2$ SCFTs with $\varkappa \neq \{1,2\}$ we must consider the extensions of any diagram in \eqref{eq:extensions_An} as well. 
We find that no new diagram appear when considering these extensions, in other words any extension of 
an extension of $A_n$ can be obtained as an extension of $A_{n+1}$.

%Furthermore, as already discussed at the beginning of this section, all diagrams at high rank contain a long $A_n$ subdiagram, therefore we can repurpose the results on the extensions of $A_n$ diagrams, discussed in this section, to simplify the analysis of the extensions of other diagrams. 

%More precisely we have shown that any extension of $A_n$ is equivalent up to conjugation to a diagram of the form:
The analysis of extensions of other diagrams is aided by the fact that high-rank diagrams in  \eqref{eq:extensions_An} contain long $A_n$ subgraphs. We showed that, given a diagram $D$ with an $A_n$ subgraph, the connections between the new node and the $A_n$ subgraph can be reduced to the following form:
\begin{equation}	\label{eq:fully_simplified_An_extension}
\diagramA[scale=1,rotate=0, newarrowstyle1/.append style={color=black}](0)(\alpha,\beta,\gamma)
\end{equation}
with:
\begin{equation}	\label{eq:possible_abg}
(\alpha,\beta,\gamma)
=
\left\{
\begin{array}{ll}
(1,0,0),\; (0,1,0),\; (0,0,1), \qquad \qquad &\varkappa\neq\{1,2\}
\\
(1+i,0,0), \;(1,i,0), \; c.c. & \varkappa \in \left\{4,\frac{3}{4} \right\}
\\
(1,\omega,0),\; (0,1,\omega), \; c.c. & \varkappa \in \left\{3,\frac{2}{3}, 6, \frac{5}{6} \right\}
\end{array}
\right.
\end{equation}
and all the changes of basis necessary to put the diagram in the form \eqref{eq:fully_simplified_An_extension} are shifts in the new charge, schematically $q_0 \to q_0 + q_i$, which do not affect the diagram $D$. In Section \ref{subsec:Jn} we exploit this to systematically analyze the extensions of $J_n$.

%These moves do not affect the arrows that are not connected to $q_0$, therefore they can be performed even when the $A_n$ diagram is a subgraph of a bigger diagram. 

\begin{comment}
As an example any extension of the $J_n$ diagram can be put in the following form:
\begin{equation}	\label{eq:Jn_simplified_extension}
\sdiagramJ[scale=1,rotate=0, colorAnsubgraph/.append style={color=orange}](0)(\alpha,\beta,\gamma,\delta)
\end{equation}
through the moves described in this section. In \eqref{eq:Jn_simplified_extension} we colored in orange the $A_{n-1}$ subdiagram considered for the simplification. The triple $(\alpha,\beta,\gamma)$ is (the complex conjugate of) one of the triples in \eqref{eq:possible_abg} with $k=3$ while $\delta = \pm 1, \pm \omega, \pm \omega^2$.
Then the amount of extensions that one must consider is significantly reduced and the analysis can be carried out on a case by case basis. 
In the next Section we apply this procedure for the case of $J_n$ diagrams. We have studied the extensions of all the other diagrams, the results are summarized in Figure \ref{fig:flowchart}.
\end{comment}

Before analyzing the extensions of other diagrams we should ask wether the extensions of $A_n$ in \eqref{eq:extensions_An} are all inequivalent. 
A simple invariant that allows us to distinguish two diagrams 
is the determinant of the associated Hermitian matrix. % because two diagrams can be equivalent only if they have the same determinant. 
The determinants for all diagrams are summarized in Table \ref{tab:results} and the only pairs of diagrams that have the same rank and determinant are $E_6$, $J_6$ and $E_7$, $B_7$:
\begin{equation}
E_6 \stackrel{?}{=} J_6
\qquad \qquad 
E_7 \stackrel{?}{=} B_7
\end{equation}
These pairs of lattices can be distinguished by their kissing number, which is the number of charges with smallest norm excluding the origin of the lattice. The $E_6$, $E_7$ and $B_7$ charge lattices are 
closely related to
%complexified versions of 
the root lattices of the corresponding Lie algebras. In particular the smallest charges are given by the roots of the Lie algebra times a phase $\zeta_k ^\ell$, with $\ell = 1, \dots, k/2-1$.
Then the $E_6$ charge lattice with $k=6$ has kissing number $72\times 3 = 216$, and the $E_7$ and $B_7$ charge lattices with $k=4$ have kissing numbers $252$ and $196$ respectively.
Finally the $J_6$ charge lattice is given by the complex root lattice of the CCRG $G(3,3,6)$ \cite{CCRG} which has kissing number $270$.
Therefore all charge lattices in Table \ref{tab:results} that only include long charges are inequivalent.
Similarly diagrams that include short charges are never equivalent to diagrams that do not include short charges. Therefore all the diagrams in Table \ref{tab:results} correspond to inequivalent charge lattices.

%%%%%%%%%%%%%%%%%%%%%%%%%%%%%%%%%%%%%%%%%%%%%%%%%%%%%%%%%%%%%%%%%%%%%%%%%%%%%%%%
\subsection{Extensions of $J_n$}
\label{subsec:Jn}

We now consider the possible extensions of the $J_n$, $n\geq 3$ diagrams and show that they are equivalent to either the $J_{n+1}$ or $K_{n+1}$ diagram. Let us start with the edge case $J_3$, whose extensions that only contain long charges are of the form:
\begin{equation}
\sdiagramJthree[scale=1,  colorAnsubgraph/.append style={color=orange}](0)(\alpha,\beta,\delta)(-1,-1,\omega)
,\qquad
\begin{array}{l}
	(\alpha,\beta) = (0,1),\; (1,0),\; (1,\omega),\; (1,\overline{\omega})
	\\
	\delta = \pm 1, \pm \omega, \pm \omega^2
\end{array}
\end{equation}
where we highlighted in orange the $A_2$ subdiagram used to simplify the extension.
The diagrams with positive determinant that do not contain any loop with weight $-\omega^2, -\overline{\omega^2}$ are\footnote{As discussed in the previous Section the presence of a loop with weight $-\omega^2, -\overline{\omega^2}$ implies that the lattice includes short charges. Unless otherwise stated in this Section we always consider lattices that only include long charges.}:
\begin{equation}	\label{eq:extensions_J3_lst}
\begin{array}{l}
\sdiagramJthree[scale=1](0)(1,-1,0)(-1,-1,\omega) ,
\qquad\quad 
\sdiagramJthree[scale=1](0)(\omega,-1,0)(-1,-1,\omega) ,
\qquad\quad 
\sdiagramJthree[scale=1](0)(0,-1,0)(-1,-1,\omega) ,
\\
\sdiagramJthree[scale=1](0)(\omega,0,-1)(-1,-1,\omega) ,
\qquad \quad
\sdiagramJthree[scale=1](0)(\omega^2,0,-1)(-1,-1,\omega) ,
\qquad \quad
\sdiagramJthree[scale=1](0)(0,0,-1)(-1,-1,\omega) ,
\end{array}
\end{equation}

The first and fourth and fifth diagrams in \eqref{eq:extensions_J3_lst} are equivalent up to permutations of the nodes, and all the diagrams in \eqref{eq:extensions_J3_lst} are equivalent to $J_4$ as can be seen from the changes of basis:

\begin{equation}
\sdiagramJthree[scale=1](0)(1,-1,0)(-1,-1,\omega) 
\quad \underrightarrow{q_1\to q_1 + q_2}\quad	
\sdiagramJthree[scale=1](0)(0,-1,0)(1,-1,\omega^2) 
\end{equation}

\begin{equation}
\sdiagramJthree[scale=1](0)(\omega,-1,0)(-1,-1,\omega)
\quad \underrightarrow{q_1\to q_1 + q_0 + q_2}\quad	
\sdiagramJthree[scale=1](0)(\omega,-1,0)(-1,-1,0)
\end{equation}

which are both equivalent up to conjugation to $J_4$.\\

We now consider the extensions of the $J_n$ diagram with $n\geq 4$,
which can be put in the following form exploiting the analysis of the extensions of $A_n$ spelled out in the previous Section:
% which can be put in the form in eq. \eqref{eq:Jn_simplified_extension}, reported here for convenience:
\begin{equation}	
\extensiondiagram[J_n]_{\alpha,\beta,\gamma,\delta} = \sdiagramJ[scale=1,rotate=0, colorAnsubgraph/.append style={color=orange}](0)(\alpha,\beta,\gamma,\delta) ,
\qquad
\end{equation}
where we highlighted in orange the $A_{n-1}$ subgraph used to simplify the extension. Here:
\begin{equation}
\begin{array}{l}
(\alpha,\beta,\gamma) = (1,0,0),\; (0,1,0), \; (0,0,1), \; (1,\omega,0) \; (0,1,\omega), \; c.c.
\\
\delta =  \pm 1, \pm \omega, \pm \omega^2
\end{array}
\end{equation}
the extensions with positive determinant that do not contain short charges are, up to conjugation, the following:
\begin{equation}	\label{eq:Jn_extension_1}
\extensiondiagram[J_n]_{\omega,0,0,-1}  = \sdiagramJ[scale=.8,rotate=0](0)(\omega,0,0,-1) ,
\qquad \forall n
\end{equation}

\begin{equation}	\label{eq:Jn_extension_2}
\extensiondiagram[J_n]_{\omega^2,0,0,-1}  = \sdiagramJ[scale=.8,rotate=0](0)(\omega^2,0,0,-1) ,
\qquad n=4,5 
\end{equation}

\begin{equation}	\label{eq:Jn_extension_3}
\extensiondiagram[J_n]_{0,0,0,-1}  = \sdiagramJ[scale=.8,rotate=0](0)(0,0,0,-1) 
,\qquad n=4,5 
\end{equation}

\begin{equation}	\label{eq:Jn_extension_4}
\extensiondiagram[J_n]_{\omega,-1,0,0}  = \sdiagramJ[scale=.8,rotate=0](0)(\omega,-1,0,0) 
,\qquad n=4,5 
\end{equation}

\begin{equation}	\label{eq:Jn_extension_5}
\extensiondiagram[J_n]_{1,-1,0,0} = \sdiagramJ[scale=.8,rotate=0](0)(1,-1,0,0) 
,\qquad n=4,5 
\end{equation}

\begin{equation}	\label{eq:Jn_extension_6}
\extensiondiagram[J_n]_{\omega,0,-1,0} = 
\sdiagramJ[scale=.8,rotate=0](0)(\omega,0,-1,0) 
,\qquad n=4,5 
\end{equation}

\begin{equation}	\label{eq:Jn_extension_7}
\extensiondiagram[J_n]_{1,0,-1,0} = 
\sdiagramJ[scale=.8,rotate=0](0)(1,0,-1,0) 
,\qquad n=4,5 
\end{equation}

\begin{equation}	\label{eq:Jn_extension_8}
\extensiondiagram[J_4]_{0,0,-1,0} = 
\vcenter{\hbox{\begin{tikzpicture}[scale=.8]

\tikzstyle{nodestyle} = [doublecirclestyle];

%nodes
\begin{scope} [every node/.style=doublecirclestyle]
	\node[nodestyle,colorAnsubgraph] (A) at (0.3,1) {1};
	\node (Adown) at (0.3,-1) {3};
	\node[nodestyle,colorAnsubgraph] (B) at (2,0) {2};
	\node[nodestyle,colorAnsubgraph] (C) at (4,0) {3};
	%new node
	\node (new) at (2,2) {0};
\end{scope}

% arrow label nodes
%\node (labelFixed) at (0,0) {$\omega$};

%edges
\begin{scope}[every node/.style={fill=white}, every edge/.style={draw,very thick}]
	\arrowedge[-1](A,B)(colorAnsubgraph)
	\arrowedge[-1](Adown,B)()
	\path[->] (A) edge node[left] {$\omega$} (Adown);
	%\arrowedge[\omega](A,Adown)()
	\arrowedge[-1](B,C)()
\end{scope}

%new edges
\begin{scope}[every node/.style={fill=white}, every edge/.style={draw,very thick}]
	\arrowedge[-1](new,C)(newarrowstyle1,bend left)
\end{scope}
\end{tikzpicture}}}
\end{equation}

The extension \eqref{eq:Jn_extension_3} is equivalent to $K_5$ or $K_6$ for $n=4,5$ respectively, while the extension \eqref{eq:Jn_extension_8} is equivalent to $J_5$. 
The other diagrams are equivalent either to $J_{n+1}$ or $K_{n+1}$ for $n=4,5$, as can be seen from the changes of basis:
\begin{equation}
\extensiondiagram[J_n]_{\omega,0,0,-1}  
\quad \underrightarrow{q_1\to q_0 + q_4}\quad	
J_{n+1}
\end{equation}

\begin{equation}
\extensiondiagram[J_n]_{\omega^2,0,0,-1}  
\quad \underrightarrow{q_1\to q_4}\quad	
K_{n+1}
\end{equation}

\begin{equation}
\extensiondiagram[J_n]_{\omega,-1,0,0}  
\quad \underrightarrow{q_2\to q_2 + q_1 + q_4+q_1}\quad	
K_{n+1}
\end{equation}

\begin{equation}
\extensiondiagram[J_n]_{1,-1,0,0} 
\quad \underrightarrow{q_2\to q_2 + q_1}\quad	
K_{n+1}
\end{equation}

\begin{equation}
\extensiondiagram[J_n]_{\omega,0,-1,0}
\quad \underrightarrow{q_3\to q_3+ q_0+q_1+q_4}\quad	
K_{n+1}
\end{equation}

\begin{equation}
\extensiondiagram[J_n]_{1,0,-1,0}
\quad \underrightarrow{ q_3\to q_3+ q_0+q_1,\; q_1 \to q_1+q_4}\quad	
K_{n+1}
\end{equation}

For the sake of readability, in the changes of basis $q_i \to q_i + \alpha q_j$ we have dropped the coefficients $\alpha$, which are uniquely fixed by requiring that the new charge $q_i+ \alpha q_j$ has length squared 2.

This concludes the analysis of $J_n$ that only include long charges. We have found that such extensions are equivalent to either $J_{n+1}$ or $K_{5,6}$:

\begin{equation}
J_n \quad \longrightarrow \quad J_{n+1}, \; K_{5,6}
\end{equation}

The extensions of other diagrams can be analyzed with the same techniques. 
We have checked that all extensions that only include long charges are equivalent to one of the diagrams with only long charges in Table \ref{tab:results}. More precisely the diagrams can be extended in the following ways:

\begin{equation}
\renewcommand\arraystretch{1.7}
\begin{array}{rl}
A_n \quad \longrightarrow \quad&\quad
\left \{
\renewcommand\arraystretch{1}
\begin{array}{ll}
A_{n+1},\; D_{n+1}, \; E_{6,7,8}, \qquad\qquad &\varkappa\neq\{1,2\}
\\
B_{n+1},\; F_{4} &\varkappa \in \left\{4,\frac{3}{4} \right\}
\\
J_{n+1}, \;K_{5,6} &\varkappa \in \left\{3,\frac{2}{3},6,\frac{5}{6} \right\}
\end{array}
\renewcommand\arraystretch{1.7}
\right.
\\
D_n \quad \longrightarrow \quad &\quad D_{n+1}, \; E_{6,7,8}
\\
E_{6,7} \quad \longrightarrow \quad &\quad E_{7,8}
\\
J_n \quad \longrightarrow \quad &\quad J_{n+1}, \; K_{5,6}
\\
K_5 \quad \longrightarrow \quad &\quad K_{6}
\\
BC_n \quad \longrightarrow \quad &\quad BC_{n+1}, \; F_{4}
\end{array}
\renewcommand\arraystretch{1}
\end{equation}
while $F_4, K_6$ and $E_8$ do not admit any extension.

%%%%%%%%%%%%%%%%%%%%%%%%%%%%%%%%%%%%%%%%%%%%%%%%%%%%%%%%%%%%%%%%%%%%%%%%%%%%%%%%
\subsection{Including short charges}
\label{subsec:short}
We now consider charge lattices involving short charges, namely charges $q_i$ with $H(q_i,q_i) = 1$. 
First we consider lattices that can be obtained as an extension by a short charge of one of the lattices with only long charges, classified in the previous Sections. We find that all these extensions are equivalent to the $\overline{A_n}$ diagram, see Table \ref{tab:results}. We then consider extensions of the  $\overline{A_n}$ diagram itself which turn out to be equivalent to $\overline{A_{n+1}}$. 
Therefore the main result of this section is that all charge lattices of irreducible $\NN=2$ SCFTs with $\varkappa\neq\{1,2\}$ that include short charges are equivalent to the $\overline{A_n}$ diagrams:
\begin{equation}
\overline{A_n} = \diagramAbar[scale=1](0)(0,0,0)
\end{equation}
\begin{equation}
\det \left( \overline{A_n} \right) = 1
\end{equation}

From the Cauchy inequality there can not be an edge between two short charges and an edge $\alpha$ between a short and a long charge must have $|\alpha| =1$:
\begin{equation}
\vcenter{\hbox{\begin{tikzpicture}
\node[singlecirclestyle] (A) at (0,0) {};
\node[doublecirclestyle] (B) at (2,0) {};
\arrowedge[\alpha](A,B)(very thick, every node/.style={fill=white})
\end{tikzpicture}}},
\qquad
\alpha = \left\{
\begin{array}{ll}
\pm1, \; \pm i \qquad\qquad & \varkappa \in \left\{4,\frac{3}{4} \right\}
\\
\pm1, \; \pm \omega, \; \pm \omega^2 \qquad\qquad & \varkappa \in \left\{3,\frac{2}{3},6,\frac{5}{6} \right\}
\end{array}
\right.
\end{equation}

As in the previous section we begin by considering extensions of the $A_n$ diagram by a new short charge $q_0$. These extensions will generally contain loops involving $q_0$, whose determinant is:
\begin{equation}
\det\left(
\vcenter{\hbox{\begin{tikzpicture}
\node[label=below:1,doublecirclestyle] (A) at (-2,0) {};
\node[label=below:$r-1$,doublecirclestyle] (B) at (2,0) {};
\node[singlecirclestyle] (top) at (0,2) {0};

\arrowedge[-1](A,B)(very thick, every node/.style={fill=white},dashed)
\arrowedge[-1](A,top)(very thick, every node/.style={fill=white})
\arrowedge[\alpha](top,B)(very thick, every node/.style={fill=white})
\end{tikzpicture}}}
\right)
= 2 + 2\Re(\alpha) -r
\end{equation}
The only loop with positive determinant is given by $\alpha=1$ and $r=3$. Furthermore the only double loop that can be built out of this single loop has vanishing determinant and is not allowed:
\begin{equation}
\det\left(
\vcenter{\hbox{\begin{tikzpicture}
\node[doublecirclestyle] (A) at (-2,0) {};
\node[doublecirclestyle] (B) at (0,0) {};
\node[doublecirclestyle] (C) at (2,0) {};
\node[singlecirclestyle] (top) at (0,2) {0};

\arrowedge[-1](A,B)(very thick, every node/.style={fill=white})
\arrowedge[-1](B,C)(very thick, every node/.style={fill=white})
\arrowedge[-1](A,top)(very thick, every node/.style={fill=white})
\arrowedge[1](top,B)(very thick, every node/.style={fill=white})
\arrowedge[-1](C,top)(very thick, every node/.style={fill=white})
\end{tikzpicture}}}
\right)
= 0
\end{equation}

Therefore the new short  charge $q_0$ can either be connected to a single node or can form a triangle loop with two adjacent nodes. 
If $q_0$ is connected to a single node $j$, then $j$ must be the leftmost or rightmost node of the $A_n$ diagram, otherwise the extension would contain the following subgraph with non-positive determinant:
\begin{equation}
\det\left(
\vcenter{\hbox{\begin{tikzpicture}
\node[doublecirclestyle] (A) at (-2,0) {};
\node[doublecirclestyle] (B) at (0,0) {};
\node[doublecirclestyle] (C) at (2,0) {};
\node[singlecirclestyle] (top) at (0,2) {0};

\arrowedge[-1](A,B)(very thick, every node/.style={fill=white})
\arrowedge[-1](B,C)(very thick, every node/.style={fill=white})
\arrowedge[1](top,B)(very thick, every node/.style={fill=white})
\end{tikzpicture}}}
\right)
= 0
\end{equation}

On the other hand if $q_0$ forms a triangle loop with two adjacent nodes, say $j$ and $j+1$, we can move the loop to the right (equivalently, left) of the $A_n$ diagram by the following change of basis:

\begin{equation}	\label{eq:triangle_move_short}
\begin{split}
\sdiagramAreducedshort[scale=1](0)(-1,1)(j,j+1)&
\\
\quad \underrightarrow{q_0\to q_0 - \sum_{\ell=j+1}^n q_\ell}\quad	&
\vcenter{\hbox{\begin{tikzpicture}
\node[label=below:1,doublecirclestyle] (A) at (-3,0) {};
\node[label=below:$n$,doublecirclestyle] (B) at (0,0) {};
\node[singlecirclestyle] (C) at (2,0) {0};
\arrowedge[-1](A,B)(very thick, every node/.style={fill=white},dashed)
\arrowedge[-1](B,C)(very thick, every node/.style={fill=white})
\end{tikzpicture}}}
\end{split}
\end{equation}
In both cases the allowed extensions of $A_n$ are equivalent to $\overline{A_{n+1}}$. 

Similarly 
one can study the extensions of the other diagrams by a short charge.
%by choosing a maximal $A_n$ subgraph and simplifying the extension using \eqref{eq:triangle_move_short}. 
We have checked that the extensions by a short charge of diagrams different from $A_n$ have non-positive determinants, and therefore are not allowed. 

Finally we must consider extensions of the $\overline{A_n}$ diagram itself. Extensions by a new short charge $q_0$ can either contain a single edge between $q_0$ and a long charge, or they can contain a triangle loop, similar to the $A_n$ extensions discussed above. By using \eqref{eq:triangle_move_short}, any allowed extension can be put in the form:
\begin{equation}
\vcenter{\hbox{\begin{tikzpicture}
\node[label=below:1,doublecirclestyle] (A) at (-3,0) {};
\node[label=below:$n$,doublecirclestyle] (B) at (0,0) {};
\node[singlecirclestyle] (C) at (2,0) {0};
\node[singlecirclestyle] (Z) at (-5,0) {0};
\arrowedge[-1](A,B)(very thick, every node/.style={fill=white},dashed)
\arrowedge[-1](B,C)(very thick, every node/.style={fill=white})
\arrowedge[-1](Z,A)(very thick, every node/.style={fill=white})
\end{tikzpicture}}}
\end{equation}
which has vanishing determinant and is not allowed. 

Now consider extensions of the $\overline{A_n}$ diagram by a long charge $q_0$, and denote as $q_n$ the only short charge of the diagram. If $q_0$ is connected to any of the long charge of $\overline{A_n}$ the extension can alternatively be thought as an extension by the short charge $q_r$, which falls into one of the cases considered above.
 %studied in Appendix \ref{app:B}. 
 On the other hand if $q_0$ is only connected to $q_r$ there is the following subgraph with non-positive determinant:
\begin{equation}
\det\left(
\vcenter{\hbox{\begin{tikzpicture}
\node[doublecirclestyle] (A) at (-2,0) {};
\node[singlecirclestyle] (B) at (0,0) {};
\node[doublecirclestyle] (C) at (2,0) {};

\arrowedge[-1](A,B)(very thick, every node/.style={fill=white})
\arrowedge[-1](B,C)(very thick, every node/.style={fill=white})
\end{tikzpicture}}}
\right)
= 0
\end{equation}
and the extension is not allowed.

Putting it all together, the only diagram which include short charges is $\overline{A_n}$, which can be obtained as an extensions of $A_{n-1}$ or $\overline{A_{n-1}}$ and can only be extended to $\overline{A_{n+1}}$.
\begin{equation}
\renewcommand\arraystretch{1.7}
\begin{array}{l}
A_n \quad \longrightarrow \quad \overline{A_{n+1}}
\\
\overline{A_{n}}\quad \longrightarrow \quad \overline{A_{n+1}}
\end{array}
\renewcommand\arraystretch{1}
\end{equation}

%%%%%%%%%%%%%%%%%%%%%%%%%%%%%%%%%%%%%%%%%%%%%%%%%%%%%%%%%%%%%%%%%%%%%%%%%%%%%%%%
%\section{1-form symmetries, non-invertible symmetries and the sporadic $K_5$ lattice}
\section{Generalized symmetries from charge lattices}
\label{sec:4}
%intro
The charge lattice of $\NN=2$ theories is closely related to the 1-form symmetry group $G^{(1)}$. The specific 1-form symmetry group $G^{(1)}$ depends in general on the choice of global structure, while the order of $G^{(1)}$ is uniquely determined by the charge lattice and can be computed as the pfaffian of the Dirac pairing.
The charges of line operators corresponding to a choice of global form are a maximal set of charges with integer Dirac pairing with the charge lattice and between themselves. 

In the theories of interest in this paper, namely $\NN=2$ SCFTs with $\varkappa \neq \{1,2\}$, the charge lattice can be identified with a complex lattice and the Dirac pairing is encoded in a positive definite Hermitian form $H$.
Then, as discussed in Section \ref{sec:2}, the order of the 1-form symmetry group is given by:
\begin{equation}
|G^{(1)}| = \det(H)
\end{equation}
The determinants for all charge lattices corresponding to irreducible theories are summarized in Table \ref{tab:results}. In particular we notice that at a fixed rank $r$ the maximum order of the 1-form symmetry group is $r+1$, realized by $\NN=4$ SYM with gauge algebra $A_n$. 
In contrast, the order of the 1-form symmetry group for a stack of theories is bounded from above by $2^r$, realized by $\NN=4$ SYM with gauge algebra $(A_1)^r$.
More precisely the 1-form symmetry group of an irreducible $\NN=2$ SCFT with $\varkappa\neq\{1,2\}$ has order:
\begin{equation}
|G^{(1)}| =
1,2,3,4,r+1
\end{equation}
where some values of $|G^{(1)}|$ only appear at specific rank for a given choice of $\varkappa$.
The 1-form symmetry group depends in general on the choice of global structure. When $|G^{(1)}|=p$ is square free the only possible 1-form symmetry group is $\zz_p$. When $|G^{(1)}|$ contains squares in its prime decomposition there are multiple choices of global structure leading to different groups which can be identified by the techniques developed in \cite{Aharony:2013hda}. In the case of interest in this paper the only diagrams where this can happen are $D_n$ and $A_n$.

%%%%%%%
\subsection{1-form symmetries of $\NN=2$ $S$-folds}
Let us consider an
%It is also interesting to notice that an 
$\NN=2$ SCFTs with $\varkappa \neq \{1,2\}$ with codimension-1 singularity supporting a theory with trivial 1-form symmetry.
The charges becoming massless at this singularity correspond to a short charge in the complex lattice.
% has a short charge, namely the states becoming massless at that singularity correspond to a short charge in the complex charge lattice.
On the other hand 
the only diagrams that include short charges are the $\overline{A_r}$ diagrams, whose determinant is 1. 
Therefore any irreducible $\NN=2$ SCFTs with $\varkappa \neq \{1,2\}$ that has a singularity supporting a rank-1 theory different from $SU(2)$ $\NN=4$ SYM has trivial 1-form symmetry.
As an example, consider the rank-$n$ MN theories, which can be engineered as the worldvolume theory on a stack of $n$ D3-branes probing an $E_{6,7,8}$ 7-brane. The Coulomb branch is the space parametrized by the positions of the D3-branes in the space transverse to both the D3 branes and the 7-brane. When only one D3-brane is on top of the 7-brane the low energy dynamic is that of the rank-1 version of the MN theory, which is known to have trivial 1-form symmetry.
Then, by the discussion above, the charge lattice of the rank-$r$ MN theories is $\overline{A_r}$, and their 1-form symmetry is trivial.
This matches with the results from the analysis of the 1-form symmetry groups of class-S theories \cite{Tachikawa:2013hya,Bhardwaj:2021pfz,Bhardwaj:2021mzl,Garding:2023unh,Albertini:2020mdx} through an alternative construction of the higher rank MN theories as class-S fixtures \cite{Benini:2009gi}.
An analogous argument shows that the $H_0$, $H_1$ and $H_2$ theories %\cite{} 
have trivial 1-form symmetry.

Similarly, consider the $\NN=2$ $S$-fold SCFTs of \cite{Apruzzi:2020pmv, Giacomelli:2020jel, Giacomelli:2020gee, Heckman:2020svr,Giacomelli:2024dbd}, which can be engineered as the worldvolume theory on a stack of $n$ D3-branes in the background of an $S$-fold and a 7-brane. The CB is parametrized by the positions of the D3-branes in the space transverse to the D3 branes (and therefore the $S$-fold) and the 7-brane.
Similarly to the case of MN theories, there is a codimension-1 singularity when a D3 is on top of the 7-brane (and therefore on top of the $S$-fold). The theory supported on this singularity is the rank-1 version of the $\NN=2$ $S$-fold SCFT, which have trivial 1-form symmetry.
%Then the charge lattice of any rank-$n$ $\NN=2$ $S$-fold SCFT is $\overline{A_n}$ and their 1-form symmetry is trivial.
Then any $\NN=2$ $S$-fold SCFT with $\varkappa\neq\{1,2\}$ has charge lattice $\overline{A_n}$ and trivial 1-form symmetry. Most $\NN=2$ $S$-folds have $\varkappa\neq\{1,2\}$, with the exception of some of the $\mathcal{T}_{D_4, 2}^{(r)}$, $\mathcal{T}_{D_4, 3}^{(r)}$ and $\mathcal{T}_{\varnothing, 6}^{(r)}$ theories \cite{Giacomelli:2020jel,Giacomelli:2020gee}.
It would be interesting to analyze the generalized symmetry of these theories further.

%K_5 analysis: 1-form sym, global structures, non-invertibles
\subsection{The sporadic $K_{5,6}$ lattices}
\label{sec:K5}
Let us consider the $K_5$ and $K_6$ lattices in more detail. 
In doing so we will also highlight some properties that can be inferred from any charge lattice in Table \ref{tab:results}.
Most of the charge lattices in our classification are realized by some known $\NN=2$ SCFT and often we know multiple SCFTs realizing the same charge lattice, as in the case of MN theories, $H_{0,1,2}$ theories and $\NN=2$ $S$-folds which all have the $\overline{A_n}$ charge lattice.
See Table \ref{tab:results} for some examples of SCFTs realizing the various charge lattices.
The only exceptions are the $K_5$ and $K_6$ lattices which, to best of our knowledge, are not realized by any known $\NN=2$ SCFT.
Therefore in this Section we analyze some of the properties that a putative SCFT realizing the $K_{5,6}$ lattices must posses.

Before delving into details we notice that the $K_5$ and $K_6$ lattices are compatible with the CB geometries $\mathbb{C}^5/G_{33}$ and $\mathbb{C}^6/G_{34}$ respectively \cite{Kaidi:2022lyo}.  These would correspond to the CB of new $\NN=3$ SCFTs, but to this date there is no known geometric engineering setup or field theoretical construction realizing these CB geometries\footnote{In \cite{Kaidi:2022lyo} two additional $\NN=3$ CB geometries where proposed, namely $\mathbb{C}^3/G_{24}$ and $\mathbb{C}^4/G_{29}$. The latter is compatible with the $F_4$ lattice, while $G_{24}$ is defined over $\zz\left[\frac{1}{2}(1+i \sqrt{7}) \right]$ and 
it does not seem to be compatible with any charge lattice in Table \ref{tab:results}. }.
The $K_{5,6}$ lattices only include short charges, therefore any codimension-1 singularity of the CB must support $\NN=4$ $SU(2)$ SYM. Furthermore their determinant, which is equal to the order of the 1-form symmetry group, is given by:
\begin{equation}
\begin{array}{ll}
\det(K_5) = 2 \quad \to \quad&\quad G^{(1)} = \zz_2
\\
\det(K_6) = 1 \quad \to \quad&\quad \text{trivial 1-form symmetry}
\end{array}
\end{equation}

The $K_5$ lattice has a non-trivial 1-form symmetry, therefore let us consider the possible global forms that an SCFT realizing this lattice can have.
The line operators $\ell$ that can be defined are those with integer Dirac pairing with the elements of the charge lattice. In the notation used in this paper this constraint reads:
\begin{equation}
\ell = (\alpha_1,\alpha_2,\alpha_3,\alpha_4,\alpha_5,\alpha_6)
\end{equation}
\begin{equation}
H_{K_5} \cdot \ell^T \in \zz[\omega]^5
\end{equation}
where $H_{K_5}$ is the Hermitian matrix associated to the $K_5$ diagram:

\begin{equation}
H_{K_5} = \left(
\begin{array}{ccccc}
 2 & 0 & -1 & \omega  & 0 \\
 0 & 2 & -1 & 0 & 0 \\
 -1 & -1 & 2 & -1 & 0 \\
 \overline{\omega}  & 0 & -1 & 2 & -1 \\
 0 & 0 & 0 & -1 & 2 \\
\end{array}
\right)
\end{equation}

We identify two line operators if they are related by the insertion of a local operator, whose charge is in $\zz[\omega]^5$:
\begin{equation}
\ell \sim \ell + \zz[\omega]^5
\end{equation}
there are four solutions to this constraint which can be parametrized as follows:
\begin{equation}	\label{eq:K5_ell_delta}
\ell_\delta = (-i \sqrt{3},\; 1,\; 2,\; 2(1+\omega),\;1+\omega)\delta,
\qquad
\delta = 0, \frac{1}{2}, \frac{\omega}{2}, \frac{1+\omega}{2}
\end{equation}
where $\delta = 0$ correspond to the trivial line operator while $\delta \neq 0$ correspond to lines whose charge is not already included in the charge lattice $K_5$.
The lines in \eqref{eq:K5_ell_delta} are not all mutually local and therefore can not be simultaneously included in the theory. Indeed the Dirac pairing between two lines parametrized by $\delta$ and $\delta'$ is:
\begin{equation}
\langle \ell_{\delta}, \ell_{\delta'} \rangle =  
%H( \ell_{\delta}, \ell_{\delta'}) =
\ell_{\delta} \cdot H_{K_5} \cdot \ell_{\delta'}^T 
=  6 \overline{\delta} \delta'
\notin \zz[\omega]
\end{equation}

%the Dirac pairing is:
%\begin{equation}
%\overline{\ell_{\delta}} \cdot H_{K_5} \cdot \ell_{\delta'}^T = 6 \overline{\delta} \delta'
%\notin \zz[\omega]
%\end{equation}

Then there are three choices for the global structure of the theory which are parametrized by the choices of maximal sets of line operators that preserve the Dirac quantization condition: %There are 3 such choices, namely:
\begin{equation}
\begin{array}{rl}
\text{I}: \;\; &\ell_{0}, \ell_{1/2}
\\
\text{II}: \;\; &\ell_{0}, \ell_{\omega/2}
\\
\text{III}: \;\; & \ell_{0}, \ell_{(1+\omega)/2}
\end{array}
\end{equation}

The $\zz_3$ (or $\zz_3 \in \zz_6$ for $\varkappa \in \left\{ 6, \frac{5}{6} \right\}$) 0-form symmetry  acts on the complex charge lattice by multiplication by $\omega^2$. The action on the lines $\ell_\delta$ is:
\begin{equation}
\zz_3: \; \ell_\delta \quad \longrightarrow \quad \ell_{\omega^2 \delta}
\end{equation}
and exchanges the three choices of global structure as follows:
\begin{equation}
\zz_3: 
\vcenter{\hbox{\begin{tikzpicture}[scale=1.5]
\node[fill=white] (I) at (0,0) {I};
\node[fill=white] (II) at (2,0) {III};
\node[fill=white] (III) at (1,-1.73) {II};
\path[->, bend left=45] (I) edge (II);
\path[->, bend left=45] (II) edge (III);
\path[->, bend left=45] (III) edge (I);
\end{tikzpicture}}}
\end{equation}

Then for any choice of global structure the $\zz_3$ is not a symmetry of the theory because it exchanges different global structures. In similar situations one can build a topological defect for the theory by combining the interface associated to the $\zz_3$ action with a half-space gauging of the 1-form symmetry. The resulting topological defect implements a non-invertible symmetry of the theory because its fusion rules do not respect a group law. 
In our case we found that a putative SCFT with $K_5$ charge lattice would include a non-invertible topological defect analogous to the triality defect of $SU(2)$ $\NN=4$.

The same procedure can be applied to any charge lattice in Table \ref{tab:results}: when the determinant is greater than 1 the 1-form symmetry is non-trivial and one can compute the possible global forms by solving the Dirac pairing constraints. Then one can consider the action of the $\zz_k$ global symmetries on the line operators. If some of the global forms are exchanged by $\zz_k$ then one can define a non-invertible topological defect for those choices of global forms. 
As an example, running this procedure for the $J_n$ and $BC_n$ lattices one can analyze the global forms and non-invertible symmetries of fluxless S$_3$-fold and S$_4$-folds respectively, matching  the results of \cite{Etheredge:2023ler,Amariti:2023hev}.
Similarly, the charge lattice of $\NN=4$ SYM is given by the lattice with the corresponding name in Table \ref{tab:results}. In the case of simply laced algebra some subgroup of S-duality becomes a symmetry of the theory at particular values of the complexified gauge coupling. This symmetry acts on the charge lattice as a multiplication by $\zeta_k$, therefore we can study its action on the possible global forms of the theory with the same techniques spelled out above. 
When a global form is not invariant under multiplication by $\zeta_k$, the $\zz_k$ symmetry must be combined with a half-space gauging of the 1-form symmetry, reproducing the non-invertible self-duality defects of $\NN=4$.
Therefore our approach provides a setting where the non-invertible topological defects of $\NN=4$ SYM and $\NN=3$ $S$-fold are on the same footing.

%%%%%%%%%%%%%%%%%%%%%%%%%%%%%%%%%%%%%%%%%%%%%%%%%%%%%%%%%%%%%%%%%%%%%%%%%%%%%%%%
\section{Conclusions}
\label{sec:conclusions}
In this paper we considered maximally strongly coupled $\NN=2$ SCFTs in four dimensions. 
We showed that electromagnetic charges of local states on a generic point of the CB are strongly constrained by the Dirac quantization condition and the requirement of reproducing rank-1 charge lattice on codimension-1 singularities.
For $\NN=2$ SCFTs with characteristic dimension $\varkappa \neq \{1,2\}$ we exploited these constraints to fully classify the possible charge lattices at any rank. The lattices for rank-$n$ SCFTs that are not stacks of lower rank theories are collected in Table \ref{tab:results} in a graphical notation introduced in Section \ref{sec:2}.

We reproduced the known charge lattices and 1-form symmetries of $\NN=3$ $S$-fold SCFTs and $\NN=4$ SYM with any gauge algebra.
Furthermore we argued that in some instances little information about the CB can uniquely identify the charge lattice of a theory.
For example if a  $\NN=2$ SCFTs with $\varkappa \neq \{1,2\}$ has a singularity supporting a rank-1 theory different from $SU(2)$ $\NN=4$ SYM the charge lattice is $\overline{A_n}$ and the 1-form symmetry is trivial.
We exploited this fact to determine the charge lattice of higher-rank MN theories and most $\NN=2$ $S$-folds and showed that they have trivial 1-form symmetry.

We found two consistent charge lattices, denoted as $K_5$ and $K_6$ throughout this paper that, to the best our knowledge, are not realized by any known SCFTs. 
These lattice are compatible with the moduli spaces $(\cc^3)^5/G_{33}$ and $(\cc^3)^6/G_{34}$, first considered in \cite{Kaidi:2022lyo}, which are compatible with $\NN=3$ supersymmetry.
We showed that an SCFT realizing the $K_5$ lattice would have a $\zz_2$ 1-form symmetry and we argued that it would posses a non-invertible topological defect analogous to the triality defect of $SU(2)$ $\NN=4$ SYM. 

% Orbi-$S$-folds
There are multiple directions our approach can be extended towards. 
In the context of $\NN=2$ SCFTs with $\varkappa \neq \{1,2\}$ our classification provides predictions for the charge lattice and 1-form symmetry group of new SCFTs. As an example \cite{Giacomelli:2024dbd} analyzed a setup that engineers a large number of $\NN=2$ SCFTs and provided an extended list of rank-3 SCFTs. For SCFTs with $\varkappa \neq \{1,2\}$ our results predicts that the charge lattice is either $A_3$, $J_3$, $BC_3$ or $\overline{A_3}$, with 1-form symmetry $\zz_4$ (or $\zz_2\times\zz_2$), $\zz_3$, $\zz_2$ or trivial, respectively. 
It would be interesting to identify the charge lattices of these theories.

%BPS states

%The set of electromagnetic charges provides the starting point for the study of BPS charged states. From our perspective, given a charge lattice in Table \ref{tab:results}, one could ask what are the possible charges that can be populated by BPS states. As we already discussed there are generally multiple SCFTs realizing the same charge lattice, therefore there can be multiple answers to this question, and analyzing the full spectrum of BPS states is challenging, in part due to the possible presence of walls of marginal stability.
%Nevertheless the knowledge of the charge lattice constrains the BPS spectrum, for example we known that the BPS states that become massless at codimension-1 singularity must sit in (a multiple of) a charge with length squared 1 or 2. 
% It would be interesting to study how much we can constrain the BPS spectrum given a choice of charge lattice.

% varkappa = 1,2
Finally a natural generalization of our procedure would involve relaxing the condition on the characteristic dimension. It would be interesting to study the charge lattices of SCFTs with $\varkappa=\{1,2\}$ from a bottom up perspective. The formalism of complex lattices and Hermitian forms used in this paper do not apply to the general case and most notably codimension-1 singularity can in general support IR-free theories, suggesting that charge lattices are less constrained in the general case. Nevertheless the Dirac quantization condition and the requirement of reproducing rank-1 charge lattices on singularities still apply. It would be interesting to analyze the consequences of these constraints in the most general case.

%%%%%%%%%%%
\section*{Acknowledgments}
%%%%%%%%%%%%%%%%%%
%
%
We are grateful to Valdo Tatitscheff for discussions.
The work of A.A. and S.R.  has been supported in part by the Italian Ministero dell'Istruzione, 
Universit\`a e Ricerca (MIUR) and  in part by Istituto Nazionale di Fisica Nucleare (INFN) through the “Gauge Theories, Strings, Supergravity” (GSS) research project.
The work of S.R. has been partially supported by the MUR-PRIN grant No. 2022NY2MXY.

\bibliographystyle{JHEP}
\bibliography{ref}

\providecommand{\href}[2]{#2}\begingroup\raggedright\begin{thebibliography}{10}

\bibitem{Argyres:2015gha}
P.~C. Argyres, M.~Lotito, Y.~L\"u and M.~Martone, \emph{{Geometric constraints
  on the space of $ \mathcal{N} $ = 2 SCFTs. Part II: construction of special
  K\"ahler geometries and RG flows}},
  \href{https://doi.org/10.1007/JHEP02(2018)002}{\emph{JHEP} {\bfseries 02}
  (2018) 002} [\href{https://arxiv.org/abs/1601.00011}{{\ttfamily
  1601.00011}}].

\bibitem{Argyres:2016xmc}
P.~Argyres, M.~Lotito, Y.~L\"u and M.~Martone, \emph{{Geometric constraints on
  the space of $ \mathcal{N}$ = 2 SCFTs. Part III: enhanced Coulomb branches
  and central charges}},
  \href{https://doi.org/10.1007/JHEP02(2018)003}{\emph{JHEP} {\bfseries 02}
  (2018) 003} [\href{https://arxiv.org/abs/1609.04404}{{\ttfamily
  1609.04404}}].

\bibitem{Argyres:2016xua}
P.~C. Argyres, M.~Lotito, Y.~L\"u and M.~Martone, \emph{{Expanding the
  landscape of $ \mathcal{N} $ = 2 rank 1 SCFTs}},
  \href{https://doi.org/10.1007/JHEP05(2016)088}{\emph{JHEP} {\bfseries 05}
  (2016) 088} [\href{https://arxiv.org/abs/1602.02764}{{\ttfamily
  1602.02764}}].

\bibitem{Argyres:2015ffa}
P.~Argyres, M.~Lotito, Y.~L\"u and M.~Martone, \emph{{Geometric constraints on
  the space of $ \mathcal{N} $ = 2 SCFTs. Part I: physical constraints on
  relevant deformations}},
  \href{https://doi.org/10.1007/JHEP02(2018)001}{\emph{JHEP} {\bfseries 02}
  (2018) 001} [\href{https://arxiv.org/abs/1505.04814}{{\ttfamily
  1505.04814}}].

\bibitem{Martone:2021ixp}
M.~Martone, \emph{{Testing our understanding of SCFTs: a catalogue of rank-2 $
  \mathcal{N} $ = 2 theories in four dimensions}},
  \href{https://doi.org/10.1007/JHEP07(2022)123}{\emph{JHEP} {\bfseries 07}
  (2022) 123} [\href{https://arxiv.org/abs/2102.02443}{{\ttfamily
  2102.02443}}].

\bibitem{Kaidi:2021tgr}
J.~Kaidi and M.~Martone, \emph{{New rank-2 Argyres-Douglas theory}},
  \href{https://doi.org/10.1103/PhysRevD.104.085004}{\emph{Phys. Rev. D}
  {\bfseries 104} (2021) 085004}
  [\href{https://arxiv.org/abs/2104.13929}{{\ttfamily 2104.13929}}].

\bibitem{Xie:2015rpa}
D.~Xie and S.-T. Yau, \emph{{4d N=2 SCFT and singularity theory Part I:
  Classification}},  \href{https://arxiv.org/abs/1510.01324}{{\ttfamily
  1510.01324}}.

\bibitem{Chen:2016bzh}
B.~Chen, D.~Xie, S.-T. Yau, S.~S.~T. Yau and H.~Zuo, \emph{{4D $\mathcal{N} =
  2$ SCFT and singularity theory. Part II: complete intersection}},
  \href{https://doi.org/10.4310/ATMP.2017.v21.n1.a2}{\emph{Adv. Theor. Math.
  Phys.} {\bfseries 21} (2017) 121}
  [\href{https://arxiv.org/abs/1604.07843}{{\ttfamily 1604.07843}}].

\bibitem{Wang:2016yha}
Y.~Wang, D.~Xie, S.~S.~T. Yau and S.-T. Yau, \emph{{$4d$ $\mathcal{N} = 2$ SCFT
  from complete intersection singularity}},
  \href{https://doi.org/10.4310/ATMP.2017.v21.n3.a6}{\emph{Adv. Theor. Math.
  Phys.} {\bfseries 21} (2017) 801}
  [\href{https://arxiv.org/abs/1606.06306}{{\ttfamily 1606.06306}}].

\bibitem{Chen:2017wkw}
B.~Chen, D.~Xie, S.~S.~T. Yau, S.-T. Yau and H.~Zuo, \emph{{4d $\mathcal{N}=2$
  SCFT and singularity theory Part III: Rigid singularity}},
  \href{https://doi.org/10.4310/ATMP.2018.v22.n8.a2}{\emph{Adv. Theor. Math.
  Phys.} {\bfseries 22} (2018) 1885}
  [\href{https://arxiv.org/abs/1712.00464}{{\ttfamily 1712.00464}}].

\bibitem{Xie:2021hxd}
D.~Xie and D.~Zhang, \emph{{Mixed Hodge structure and $\mathcal{N}=2$ Coulomb
  branch solution}},  \href{https://arxiv.org/abs/2107.11180}{{\ttfamily
  2107.11180}}.

\bibitem{Cecotti:2021ouq}
S.~Cecotti, M.~Del~Zotto, M.~Martone and R.~Moscrop, \emph{{The Characteristic
  Dimension of Four-Dimensional ${\mathcal {N}}$~=~2 SCFTs}},
  \href{https://doi.org/10.1007/s00220-022-04605-5}{\emph{Commun. Math. Phys.}
  {\bfseries 400} (2023) 519}
  [\href{https://arxiv.org/abs/2108.10884}{{\ttfamily 2108.10884}}].

\bibitem{Martone:2020nsy}
M.~Martone, \emph{{Towards the classification of rank-r$ \mathcal{N} $ = 2
  SCFTs. Part I. Twisted partition function and central charge formulae}},
  \href{https://doi.org/10.1007/JHEP12(2020)021}{\emph{JHEP} {\bfseries 12}
  (2020) 021} [\href{https://arxiv.org/abs/2006.16255}{{\ttfamily
  2006.16255}}].

\bibitem{Argyres:2020wmq}
P.~C. Argyres and M.~Martone, \emph{{Towards a classification of rank r$
  \mathcal{N} $ = 2 SCFTs. Part II. Special Kahler stratification of the
  Coulomb branch}}, \href{https://doi.org/10.1007/JHEP12(2020)022}{\emph{JHEP}
  {\bfseries 12} (2020) 022}
  [\href{https://arxiv.org/abs/2007.00012}{{\ttfamily 2007.00012}}].

\bibitem{Martone:2021drm}
M.~Martone and G.~Zafrir, \emph{{On the compactification of 5d theories to
  4d}}, \href{https://doi.org/10.1007/JHEP08(2021)017}{\emph{JHEP} {\bfseries
  08} (2021) 017} [\href{https://arxiv.org/abs/2106.00686}{{\ttfamily
  2106.00686}}].

\bibitem{Argyres:2020nrr}
P.~Argyres and M.~Martone, \emph{{Construction and classification of Coulomb
  branch geometries}},  \href{https://arxiv.org/abs/2003.04954}{{\ttfamily
  2003.04954}}.

\bibitem{Caorsi:2018zsq}
M.~Caorsi and S.~Cecotti, \emph{{Geometric classification of 4d $\mathcal{N}=2$
  SCFTs}}, \href{https://doi.org/10.1007/JHEP07(2018)138}{\emph{JHEP}
  {\bfseries 07} (2018) 138}
  [\href{https://arxiv.org/abs/1801.04542}{{\ttfamily 1801.04542}}].

\bibitem{Argyres:2018zay}
P.~C. Argyres, C.~Long and M.~Martone, \emph{{The Singularity Structure of
  Scale-Invariant Rank-2 Coulomb Branches}},
  \href{https://doi.org/10.1007/JHEP05(2018)086}{\emph{JHEP} {\bfseries 05}
  (2018) 086} [\href{https://arxiv.org/abs/1801.01122}{{\ttfamily
  1801.01122}}].

\bibitem{Caorsi:2018ahl}
M.~Caorsi and S.~Cecotti, \emph{{Special Arithmetic of Flavor}},
  \href{https://doi.org/10.1007/JHEP08(2018)057}{\emph{JHEP} {\bfseries 08}
  (2018) 057} [\href{https://arxiv.org/abs/1803.00531}{{\ttfamily
  1803.00531}}].

\bibitem{Caorsi:2019vex}
M.~Caorsi and S.~Cecotti, \emph{{Homological classification of 4d $ \mathcal{N}
  $ = 2 QFT. Rank-1 revisited}},
  \href{https://doi.org/10.1007/JHEP10(2019)013}{\emph{JHEP} {\bfseries 10}
  (2019) 013} [\href{https://arxiv.org/abs/1906.03912}{{\ttfamily
  1906.03912}}].

\bibitem{Argyres:2019yyb}
P.~C. Argyres, A.~Bourget and M.~Martone, \emph{{On the moduli spaces of 4d
  $\mathcal{N} = 3$ SCFTs I: triple special K\"ahler structure}},
  \href{https://arxiv.org/abs/1912.04926}{{\ttfamily 1912.04926}}.

\bibitem{Argyres:2018wxu}
P.~C. Argyres and M.~Martone, \emph{{Coulomb branches with complex
  singularities}}, \href{https://doi.org/10.1007/JHEP06(2018)045}{\emph{JHEP}
  {\bfseries 06} (2018) 045}
  [\href{https://arxiv.org/abs/1804.03152}{{\ttfamily 1804.03152}}].

\bibitem{Bourget:2018ond}
A.~Bourget, A.~Pini and D.~Rodr\'\i{}guez-G\'omez, \emph{{Gauge theories from
  principally extended disconnected gauge groups}},
  \href{https://doi.org/10.1016/j.nuclphysb.2019.02.004}{\emph{Nucl. Phys. B}
  {\bfseries 940} (2019) 351}
  [\href{https://arxiv.org/abs/1804.01108}{{\ttfamily 1804.01108}}].

\bibitem{Argyres:2019ngz}
P.~C. Argyres, A.~Bourget and M.~Martone, \emph{{Classification of all
  $\mathcal{N}\geq 3$ moduli space orbifold geometries at rank 2}},
  \href{https://doi.org/10.21468/SciPostPhys.9.6.083}{\emph{SciPost Phys.}
  {\bfseries 9} (2020) 083} [\href{https://arxiv.org/abs/1904.10969}{{\ttfamily
  1904.10969}}].

\bibitem{Argyres:2017tmj}
P.~C. Argyres, Y.~L\"u and M.~Martone, \emph{{Seiberg-Witten geometries for
  Coulomb branch chiral rings which are not freely generated}},
  \href{https://doi.org/10.1007/JHEP06(2017)144}{\emph{JHEP} {\bfseries 06}
  (2017) 144} [\href{https://arxiv.org/abs/1704.05110}{{\ttfamily
  1704.05110}}].

\bibitem{Nishinaka:2016hbw}
T.~Nishinaka and Y.~Tachikawa, \emph{{On 4d rank-one $ \mathcal{N}=3 $
  superconformal field theories}},
  \href{https://doi.org/10.1007/JHEP09(2016)116}{\emph{JHEP} {\bfseries 09}
  (2016) 116} [\href{https://arxiv.org/abs/1602.01503}{{\ttfamily
  1602.01503}}].

\bibitem{Argyres:2023eij}
P.~C. Argyres, M.~Martone and Z.~Yu, \emph{{Genus 2 Seiberg-Witten curves for
  rank 2 N=4 superYang-Mills theories}},
  \href{https://arxiv.org/abs/2312.15014}{{\ttfamily 2312.15014}}.

\bibitem{Closset:2023pmc}
C.~Closset and H.~Magureanu, \emph{{Reading between the rational sections:
  Global structures of 4d $\mathcal{N}=2$ KK theories}},
  \href{https://arxiv.org/abs/2308.10225}{{\ttfamily 2308.10225}}.

\bibitem{Furrer:2024zzu}
E.~Furrer and H.~Magureanu, \emph{{Coulomb branch surgery: Holonomy saddles,
  S-folds and discrete symmetry gaugings}},
  \href{https://arxiv.org/abs/2404.02955}{{\ttfamily 2404.02955}}.

\bibitem{Argyres:2022lah}
P.~C. Argyres and M.~Martone, \emph{{The rank 2 classification problem I: scale
  invariant geometries}},  \href{https://arxiv.org/abs/2209.09248}{{\ttfamily
  2209.09248}}.

\bibitem{Argyres:2022puv}
P.~C. Argyres and M.~Martone, \emph{{The rank 2 classification problem II:
  mapping scale-invariant solutions to SCFTs}},
  \href{https://arxiv.org/abs/2209.09911}{{\ttfamily 2209.09911}}.

\bibitem{Argyres:2022fwy}
P.~C. Argyres and M.~Martone, \emph{{The rank-2 classification problem III:
  curves with additional automorphisms}},
  \href{https://arxiv.org/abs/2209.10555}{{\ttfamily 2209.10555}}.

\bibitem{Garcia-Etxebarria:2015wns}
I.~Garc\'\i{}a-Etxebarria and D.~Regalado, \emph{{$ \mathcal{N}=3 $ four
  dimensional field theories}},
  \href{https://doi.org/10.1007/JHEP03(2016)083}{\emph{JHEP} {\bfseries 03}
  (2016) 083} [\href{https://arxiv.org/abs/1512.06434}{{\ttfamily
  1512.06434}}].

\bibitem{Garcia-Etxebarria:2016erx}
I.~Garc\'\i{}a-Etxebarria and D.~Regalado, \emph{{Exceptional $ \mathcal{N}=3 $
  theories}}, \href{https://doi.org/10.1007/JHEP12(2017)042}{\emph{JHEP}
  {\bfseries 12} (2017) 042}
  [\href{https://arxiv.org/abs/1611.05769}{{\ttfamily 1611.05769}}].

\bibitem{Apruzzi:2020pmv}
F.~Apruzzi, S.~Giacomelli and S.~Sch\"afer-Nameki, \emph{{4d $\mathcal{N}=2$
  S-folds}}, \href{https://doi.org/10.1103/PhysRevD.101.106008}{\emph{Phys.
  Rev. D} {\bfseries 101} (2020) 106008}
  [\href{https://arxiv.org/abs/2001.00533}{{\ttfamily 2001.00533}}].

\bibitem{Giacomelli:2020jel}
S.~Giacomelli, C.~Meneghelli and W.~Peelaers, \emph{{New $ \mathcal{N} $ = 2
  superconformal field theories from $ \mathcal{S} $-folds}},
  \href{https://doi.org/10.1007/JHEP01(2021)022}{\emph{JHEP} {\bfseries 01}
  (2021) 022} [\href{https://arxiv.org/abs/2007.00647}{{\ttfamily
  2007.00647}}].

\bibitem{Giacomelli:2020gee}
S.~Giacomelli, M.~Martone, Y.~Tachikawa and G.~Zafrir, \emph{{More on
  $\mathcal{N} =2$ S-folds}},
  \href{https://doi.org/10.1007/JHEP01(2021)054}{\emph{JHEP} {\bfseries 01}
  (2021) 054} [\href{https://arxiv.org/abs/2010.03943}{{\ttfamily
  2010.03943}}].

\bibitem{Heckman:2020svr}
J.~J. Heckman, C.~Lawrie, T.~B. Rochais, H.~Y. Zhang and G.~Zoccarato,
  \emph{{$S$-folds, string junctions, and $\mathcal{N} = 2$ SCFTs}},
  \href{https://doi.org/10.1103/PhysRevD.103.086013}{\emph{Phys. Rev. D}
  {\bfseries 103} (2021) 086013}
  [\href{https://arxiv.org/abs/2009.10090}{{\ttfamily 2009.10090}}].

\bibitem{Giacomelli:2024dbd}
S.~Giacomelli, R.~Savelli and G.~Zoccarato, \emph{{$\mathcal{N} = 2$
  Orbi-S-Folds}},  \href{https://arxiv.org/abs/2405.00101}{{\ttfamily
  2405.00101}}.

\bibitem{Giacomelli:2023qyc}
S.~Giacomelli and R.~Savelli, \emph{{\ensuremath{\mathscr{N}} = 1 SCFTs from
  F-theory on Orbifolds}},
  \href{https://doi.org/10.1007/JHEP08(2023)129}{\emph{JHEP} {\bfseries 08}
  (2023) 129} [\href{https://arxiv.org/abs/2304.11148}{{\ttfamily
  2304.11148}}].

\bibitem{Aharony:2016kai}
O.~Aharony and Y.~Tachikawa, \emph{{S-folds and 4d N=3 superconformal field
  theories}}, \href{https://doi.org/10.1007/JHEP06(2016)044}{\emph{JHEP}
  {\bfseries 06} (2016) 044}
  [\href{https://arxiv.org/abs/1602.08638}{{\ttfamily 1602.08638}}].

\bibitem{Tachikawa:2019dvq}
Y.~Tachikawa and G.~Zafrir, \emph{{Reflection groups and 3d $\mathcal{N}\ge $ 6
  SCFTs}}, \href{https://doi.org/10.1007/JHEP12(2019)176}{\emph{JHEP}
  {\bfseries 12} (2019) 176}
  [\href{https://arxiv.org/abs/1908.03346}{{\ttfamily 1908.03346}}].

\bibitem{Deb:2024zay}
A.~Deb and G.~Zafrir, \emph{{$\mathcal{N}=5$ SCFTs and quaternionic reflection
  groups}},  \href{https://arxiv.org/abs/2403.03971}{{\ttfamily 2403.03971}}.

\bibitem{CCRG}
G.~I. Lehrer and D.~E. E.~Taylor, \emph{{Unitary reflection groups}}, vol.~20.
  Cambridge University Press, 2009.

\bibitem{Kaidi:2022lyo}
J.~Kaidi, M.~Martone and G.~Zafrir, \emph{{Exceptional moduli spaces for
  exceptional $ \mathcal{N} $ = 3 theories}},
  \href{https://doi.org/10.1007/JHEP08(2022)264}{\emph{JHEP} {\bfseries 08}
  (2022) 264} [\href{https://arxiv.org/abs/2203.04972}{{\ttfamily
  2203.04972}}].

\bibitem{Amariti:2023qdq}
A.~Amariti and S.~Rota, \emph{{Exceptional S-fold SCFTs are almost trivial}},
  \href{https://arxiv.org/abs/2312.16608}{{\ttfamily 2312.16608}}.

\bibitem{DelZotto:2022ras}
M.~Del~Zotto and I.~n. Garc\'\i{}a~Etxebarria, \emph{{Global structures from
  the infrared}}, \href{https://doi.org/10.1007/JHEP11(2023)058}{\emph{JHEP}
  {\bfseries 11} (2023) 058}
  [\href{https://arxiv.org/abs/2204.06495}{{\ttfamily 2204.06495}}].

\bibitem{Argyres:2022kon}
P.~C. Argyres, M.~Martone and M.~Ray, \emph{{Dirac pairings, one-form
  symmetries and Seiberg-Witten geometries}},
  \href{https://doi.org/10.1007/JHEP09(2022)020}{\emph{JHEP} {\bfseries 09}
  (2022) 020} [\href{https://arxiv.org/abs/2204.09682}{{\ttfamily
  2204.09682}}].

\bibitem{Gaiotto:2010be}
D.~Gaiotto, G.~W. Moore and A.~Neitzke, \emph{{Framed BPS States}},
  \href{https://doi.org/10.4310/ATMP.2013.v17.n2.a1}{\emph{Adv. Theor. Math.
  Phys.} {\bfseries 17} (2013) 241}
  [\href{https://arxiv.org/abs/1006.0146}{{\ttfamily 1006.0146}}].

\bibitem{Gaiotto:2014kfa}
D.~Gaiotto, A.~Kapustin, N.~Seiberg and B.~Willett, \emph{{Generalized Global
  Symmetries}}, \href{https://doi.org/10.1007/JHEP02(2015)172}{\emph{JHEP}
  {\bfseries 02} (2015) 172} [\href{https://arxiv.org/abs/1412.5148}{{\ttfamily
  1412.5148}}].

\bibitem{Aharony:2013hda}
O.~Aharony, N.~Seiberg and Y.~Tachikawa, \emph{{Reading between the lines of
  four-dimensional gauge theories}},
  \href{https://doi.org/10.1007/JHEP08(2013)115}{\emph{JHEP} {\bfseries 08}
  (2013) 115} [\href{https://arxiv.org/abs/1305.0318}{{\ttfamily 1305.0318}}].

\bibitem{Amariti:2023hev}
A.~Amariti, D.~Morgante, A.~Pasternak, S.~Rota and V.~Tatitscheff,
  \emph{{One-form symmetries in $\mathcal{N} = 3$ S-folds}},
  \href{https://doi.org/10.21468/SciPostPhys.15.4.132}{\emph{SciPost Phys.}
  {\bfseries 15} (2023) 132}
  [\href{https://arxiv.org/abs/2303.07299}{{\ttfamily 2303.07299}}].

\bibitem{Garding:2023unh}
E.~R. G\r{a}rding, \emph{{Defect groups of class $\mathcal{S}$ theories from
  the Coulomb branch}},  \href{https://arxiv.org/abs/2311.16224}{{\ttfamily
  2311.16224}}.

\bibitem{Imamura:2016udl}
Y.~Imamura, H.~Kato and D.~Yokoyama, \emph{{Supersymmetry Enhancement and
  Junctions in S-folds}},
  \href{https://doi.org/10.1007/JHEP10(2016)150}{\emph{JHEP} {\bfseries 10}
  (2016) 150} [\href{https://arxiv.org/abs/1606.07186}{{\ttfamily
  1606.07186}}].

\bibitem{Agarwal:2016rvx}
P.~Agarwal and A.~Amariti, \emph{{Notes on S-folds and $ \mathcal{N} $ = 3
  theories}}, \href{https://doi.org/10.1007/JHEP09(2016)032}{\emph{JHEP}
  {\bfseries 09} (2016) 032}
  [\href{https://arxiv.org/abs/1607.00313}{{\ttfamily 1607.00313}}].

\bibitem{Choi:2021kmx}
Y.~Choi, C.~Cordova, P.-S. Hsin, H.~T. Lam and S.-H. Shao, \emph{{Noninvertible
  duality defects in 3+1 dimensions}},
  \href{https://doi.org/10.1103/PhysRevD.105.125016}{\emph{Phys. Rev. D}
  {\bfseries 105} (2022) 125016}
  [\href{https://arxiv.org/abs/2111.01139}{{\ttfamily 2111.01139}}].

\bibitem{Kaidi:2021xfk}
J.~Kaidi, K.~Ohmori and Y.~Zheng, \emph{{Kramers-Wannier-like Duality Defects
  in (3+1)D Gauge Theories}},
  \href{https://doi.org/10.1103/PhysRevLett.128.111601}{\emph{Phys. Rev. Lett.}
  {\bfseries 128} (2022) 111601}
  [\href{https://arxiv.org/abs/2111.01141}{{\ttfamily 2111.01141}}].

\bibitem{Bhardwaj:2022yxj}
L.~Bhardwaj, L.~E. Bottini, S.~Schafer-Nameki and A.~Tiwari,
  \emph{{Non-Invertible Higher-Categorical Symmetries}},
  \href{https://doi.org/10.21468/SciPostPhys.14.1.007}{\emph{SciPost Phys.}
  {\bfseries 14} (2023) 007}
  [\href{https://arxiv.org/abs/2204.06564}{{\ttfamily 2204.06564}}].

\bibitem{Choi:2022zal}
Y.~Choi, C.~Cordova, P.-S. Hsin, H.~T. Lam and S.-H. Shao,
  \emph{{Non-invertible Condensation, Duality, and Triality Defects in 3+1
  Dimensions}},  \href{https://arxiv.org/abs/2204.09025}{{\ttfamily
  2204.09025}}.

\bibitem{Kaidi:2022uux}
J.~Kaidi, G.~Zafrir and Y.~Zheng, \emph{{Non-invertible symmetries of $
  \mathcal{N} $ = 4 SYM and twisted compactification}},
  \href{https://doi.org/10.1007/JHEP08(2022)053}{\emph{JHEP} {\bfseries 08}
  (2022) 053} [\href{https://arxiv.org/abs/2205.01104}{{\ttfamily
  2205.01104}}].

\bibitem{Bhardwaj:2022lsg}
L.~Bhardwaj, S.~Schafer-Nameki and J.~Wu, \emph{{Universal Non-Invertible
  Symmetries}}, \href{https://doi.org/10.1002/prop.202200143}{\emph{Fortsch.
  Phys.} {\bfseries 70} (2022) 2200143}
  [\href{https://arxiv.org/abs/2208.05973}{{\ttfamily 2208.05973}}].

\bibitem{Bartsch:2022mpm}
T.~Bartsch, M.~Bullimore, A.~E.~V. Ferrari and J.~Pearson,
  \emph{{Non-invertible Symmetries and Higher Representation Theory I}},
  \href{https://arxiv.org/abs/2208.05993}{{\ttfamily 2208.05993}}.

\bibitem{Antinucci:2022vyk}
A.~Antinucci, F.~Benini, C.~Copetti, G.~Galati and G.~Rizi, \emph{{The
  holography of non-invertible self-duality symmetries}},
  \href{https://arxiv.org/abs/2210.09146}{{\ttfamily 2210.09146}}.

\bibitem{Bhardwaj:2022kot}
L.~Bhardwaj, S.~Schafer-Nameki and A.~Tiwari, \emph{{Unifying Constructions of
  Non-Invertible Symmetries}},
  \href{https://arxiv.org/abs/2212.06159}{{\ttfamily 2212.06159}}.

\bibitem{Bartsch:2022ytj}
T.~Bartsch, M.~Bullimore, A.~E.~V. Ferrari and J.~Pearson,
  \emph{{Non-invertible Symmetries and Higher Representation Theory II}},
  \href{https://arxiv.org/abs/2212.07393}{{\ttfamily 2212.07393}}.

\bibitem{Bhardwaj:2022maz}
L.~Bhardwaj, L.~E. Bottini, S.~Schafer-Nameki and A.~Tiwari,
  \emph{{Non-Invertible Symmetry Webs}},
  \href{https://arxiv.org/abs/2212.06842}{{\ttfamily 2212.06842}}.

\bibitem{Heckman:2022xgu}
J.~J. Heckman, M.~Hubner, E.~Torres, X.~Yu and H.~Y. Zhang, \emph{{Top Down
  Approach to Topological Duality Defects}},
  \href{https://arxiv.org/abs/2212.09743}{{\ttfamily 2212.09743}}.

\bibitem{Nardoni:2024sos}
E.~Nardoni, M.~Sacchi, O.~Sela, G.~Zafrir and Y.~Zheng, \emph{{Dimensionally
  Reducing Generalized Symmetries from (3+1)-Dimensions}},
  \href{https://arxiv.org/abs/2403.15995}{{\ttfamily 2403.15995}}.

\bibitem{DelZotto:2024tae}
M.~Del~Zotto, S.~N. Meynet and R.~Moscrop, \emph{{Remarks on Geometric
  Engineering, Symmetry TFTs and Anomalies}},
  \href{https://arxiv.org/abs/2402.18646}{{\ttfamily 2402.18646}}.

\bibitem{Bhardwaj:2023kri}
L.~Bhardwaj, L.~E. Bottini, L.~Fraser-Taliente, L.~Gladden, D.~S.~W. Gould,
  A.~Platschorre et~al., \emph{{Lectures on generalized symmetries}},
  \href{https://doi.org/10.1016/j.physrep.2023.11.002}{\emph{Phys. Rept.}
  {\bfseries 1051} (2024) 1}
  [\href{https://arxiv.org/abs/2307.07547}{{\ttfamily 2307.07547}}].

\bibitem{Cohen1976}
A.~M. Cohen, \emph{Finite complex reflection groups}, {\emph{Annales
  scientifiques de l'École Normale Supérieure} {\bfseries 9} (1976) 379}.

\bibitem{Tachikawa:2013hya}
Y.~Tachikawa, \emph{{On the 6d origin of discrete additional data of 4d gauge
  theories}}, \href{https://doi.org/10.1007/JHEP05(2014)020}{\emph{JHEP}
  {\bfseries 05} (2014) 020} [\href{https://arxiv.org/abs/1309.0697}{{\ttfamily
  1309.0697}}].

\bibitem{Bhardwaj:2021pfz}
L.~Bhardwaj, M.~Hubner and S.~Schafer-Nameki, \emph{{1-form Symmetries of 4d
  N=2 Class S Theories}},
  \href{https://doi.org/10.21468/SciPostPhys.11.5.096}{\emph{SciPost Phys.}
  {\bfseries 11} (2021) 096}
  [\href{https://arxiv.org/abs/2102.01693}{{\ttfamily 2102.01693}}].

\bibitem{Bhardwaj:2021mzl}
L.~Bhardwaj, S.~Giacomelli, M.~H\"ubner and S.~Sch\"afer-Nameki,
  \emph{{Relative defects in relative theories: Trapped higher-form symmetries
  and irregular punctures in class S}},
  \href{https://doi.org/10.21468/SciPostPhys.13.4.101}{\emph{SciPost Phys.}
  {\bfseries 13} (2022) 101}
  [\href{https://arxiv.org/abs/2201.00018}{{\ttfamily 2201.00018}}].

\bibitem{Albertini:2020mdx}
F.~Albertini, M.~Del~Zotto, I.~n. Garc\'\i{}a~Etxebarria and S.~S. Hosseini,
  \emph{{Higher Form Symmetries and M-theory}},
  \href{https://doi.org/10.1007/JHEP12(2020)203}{\emph{JHEP} {\bfseries 12}
  (2020) 203} [\href{https://arxiv.org/abs/2005.12831}{{\ttfamily
  2005.12831}}].

\bibitem{Benini:2009gi}
F.~Benini, S.~Benvenuti and Y.~Tachikawa, \emph{{Webs of five-branes and N=2
  superconformal field theories}},
  \href{https://doi.org/10.1088/1126-6708/2009/09/052}{\emph{JHEP} {\bfseries
  09} (2009) 052} [\href{https://arxiv.org/abs/0906.0359}{{\ttfamily
  0906.0359}}].

\bibitem{Etheredge:2023ler}
M.~Etheredge, I.~Garcia~Etxebarria, B.~Heidenreich and S.~Rauch, \emph{{Branes
  and symmetries for $\mathcal N=3$ S-folds}},
  \href{https://arxiv.org/abs/2302.14068}{{\ttfamily 2302.14068}}.

\end{thebibliography}\endgroup

\end{document}